\documentclass[10pt,a4paper]{iopart}
\usepackage{iopams}  
\usepackage{graphicx,psfrag,bbm,latexsym,color,dcolumn,bm,dsfont,bbm,color,
mathrsfs,bbold,latexsym,amsfonts,amssymb}

\newcommand{\media}[1]{\int d\mu_b~ #1}

\begin{document}


\title[Ising spin glass models versus Ising models: III]
{Ising spin glass models versus Ising models: \\
an effective mapping at high temperature III.
Rigorous formulation and detailed proof for \\ 
general graphs}

\author{Massimo Ostilli$^{1,2}$}
\address{$^1$\ Departamento de Fisica,
Universidade de Aveiro,
Campus Universitario de Santiago 3810-193, Aveiro, Portugal.}
\address{$^2$\ Center for Statistical Mechanics and Complexity, Istituto 
Nazionale per la Fisica della Materia, 
Unit\`a di Roma 1, Roma 00185, Italy.}

\ead{massimo.ostilli@roma1.infn.it}

\date{\today}

\begin{abstract}
Recently, it has been shown that, when the dimension of a graph
turns out to be infinite dimensional in a broad sense,
the upper critical surface and the corresponding critical behavior 
of an arbitrary Ising spin glass model defined over
such a graph, can be exactly mapped on the critical surface and behavior
of a non random Ising model. 
A graph can be infinite dimensional in a strict sense, like the
fully connected graph, or in a broad sense, as happens on a Bethe
lattice and in many random graphs. 
In this paper, we firstly introduce our definition of dimensionality
which is compared to the standard definition and readily
applied to test the infinite dimensionality of a large class of graphs which,
remarkably enough, includes even graphs where the tree-like approximation
(or, in other words, the Bethe-Peierls approach), in general, may be wrong.
Then, we derive a detailed proof of the mapping for all the graphs
satisfying this condition.
As a byproduct, the mapping provides immediately a very general Nishimori law.
\end{abstract}


\maketitle

\section{Introduction}

In a recent work \cite{MOI}, we have shown that, when the dimension 
of the graph over which an Ising spin glass model is defined turns out 
to be infinite dimensional, 
the critical surface and the critical behavior of the model
between the paramagnetic (P) and the disordered ferro/antiferromagnetic 
(F/AF) or the spin glass (SG) phases in the P regions, for shortness 
\textit{the upper critical surface and the corresponding critical behavior},
can be determined through a simple mapping with a \textit{related Ising model};
a non random model.
Infinite dimensional here 
includes two families of graphs: the ones in which the number of first
neighbors goes to infinity, as referred to as 
\textit{infinite dimensional in the strict sense}, 
and the ones characterized by the fact that the probability $p(l)$
that two randomly chosen infinitely long paths overlap for $l$
bonds, goes to zero sufficiently fast for $l$ going to infinity, 
as referred to as 
\textit{infinite dimensional in the broad sense}.
As examples, the first class includes the fully connected graph,
over which one defines the Sherrington-Kirkpatrick model \cite{SK,Parisi},
whereas the second class includes models defined over Bethe lattices
and random graphs.

In the Ref. \cite{MOI} we have derived the mapping for the models
infinite dimensional in the strict sense, whereas, except for the Bethe
lattice case, we have only provided plausible arguments 
for the other class of models to which we have
already applied the mapping in \cite{MOII}.

In this paper we give a complete proof of the mapping for the models
infinite dimensional in the broad sense providing a sufficient
condition on $p(l)$ for the mapping to become exact. 
We shall show that this condition requires that
$p(l)$ decays exponentially fast for $l$ going to infinity, Eq. (\ref{PROB2}).
We will see in fact that many graphs of interest, 
including all the ones considered in the Ref. \cite{MOII}, 
satisfy this condition.

In spite of the great simplicity of the equations of the mapping 
(\ref{GI}-\ref{mapp01g}), 
their rigorous proof is quite far from being simple. 
In fact, even within the context of the replica trick, remarkable efforts, 
involving the use of probability theory and functional analysis, have been required in deriving
the proof. We stress however that our necessity for having a complete proof of the mapping is not 
due only to a mathematical exigence, but to an urgent and practical motivation.
In fact, although in the previous papers \cite{MOI} and \cite{MOII} we have checked the mapping 
on a number of different cases, all those cases (except the
Sherrington-Kirkpatrick model, which is infinite dimensional in the strict
sense) belong to a class of models whose graphs,
roughly speaking, are characterized for having a finite number of closed paths per vertex. 
For this class of models, one might suspect that the mapping works because of the tree-like
approximation, the loops here being in a sense negligible.
However, there exists another class of models whose graphs have instead an infinite number of closed
paths per vertex but the overlap between two arbitrarily chosen paths remains sufficiently small
so that $p(l)$ decays exponentially in $l$ and the mapping remains exact. 
We point out that in this latter class of models, 
applying the tree-like approximation 
and neglecting correlations due to loops (in other words
using the Bethe-Peierls approximation \cite{NOTA}), in general, 
may lead to wrong results.

After introducing the models in Sec. 2, in Sec. 3 we provide the definition of
infinite dimensionality in the broad sense showing its connection with the
standard definition of dimensionality. In Sec. 4 we provide a short list of
graphs which are infinite dimensional in the broad sense. The mapping and its
proof are given in the Secs. 5-9. In Sec. 10 we show that the mapping leads to a
general Nishimori law. Finally, in Sec. 11 some conclusions and outlooks are drawn.

\section{Models}

\label{model}
Let be given a graph $g$ of $N$ vertices.
The set of links $\Gamma$ will be
defined through the adjacency matrix of the graph, $g_{i,j}=0,1$:
\begin{eqnarray}
\label{Gamma1}
\Gamma\equiv \{b=\left(i_b,i_b\right): i_b,j_b \in g,
~ g_{i_b,j_b}=1,~ i_b<j_b\}.
\end{eqnarray} 
The set of links of the 
fully connected graph will be indicated with $\Gamma_f$:
\begin{eqnarray}
\label{Gammaf}
\Gamma_f\equiv \{b=\left(i_b,i_b\right): i_b,j_b=1,\ldots,N, ~ i_b<j_b\}.
\end{eqnarray} 

The Hamiltonian of the spin glass with two-body interactions can be written as
\begin{eqnarray}
\label{H}
H\left(\{\sigma_i\};\{J_b\};\{h_i\}\right)
\equiv -\sum_{b\in\Gamma} J_b \tilde{\sigma}_b+\sum_{i=1}^N h_i \sigma_i,
\end{eqnarray} 
where 
the $h_i$'s are arbitrary 
external fields, the $J_b$'s are 
quenched couplings, $\sigma_i$ is an Ising variable at the site $i$, and 
$\tilde{\sigma}_b$ stays for the product
of two Ising variables, $\tilde{\sigma}_b=\sigma_{i_{b}}\sigma_{j_{b}}$, with
$i_{b}$ and $j_{b}$ such that $b=\left(i_b,j_b\right)$. 

The free energy $F$ is defined by
\begin{eqnarray}
\label{logZ}
-\beta F\equiv \int d\mathcal{P}\left(\{J_b\}\right)
\log\left(Z\left(\{J_b\};\{h_i\}\right)\right),
\end{eqnarray} 
where $Z\left(\{J_b\};\{h_i\}\right)$ 
is the partition function of the quenched system
\begin{eqnarray}
\label{Z}
Z\left(\{J_b\};\{h_i\}\right)= \sum_{\{\sigma_b\}}e^{-\beta 
H\left(\{\sigma_i\};\{J_b\};\{h_i\}\right)}, 
\end{eqnarray} 
and $d\mathcal{P}\left(\{J_b\}\right)$ 
is a product measure over all the possible bonds $b$ given 
in terms of normalized measures $d\mu_b\geq 0$ 
(we are considering a general measure $d\mu_b$ 
allowing also for a possible dependence on the bonds) 
\begin{eqnarray}
\label{dP}
d\mathcal{P}\left(\{J_b\}\right)\equiv \prod_{b\in\Gamma_f} 
d\mu_b\left( J_b \right),
\quad \int d\mu_b\left( J_b \right) =1.
\end{eqnarray}

We will take the Boltzmann constant $K_B=1$. 
A generic inverse critical temperature of the spin glass model, if any, 
will be indicated 
with $\beta_c$;
finally the density free energy in the thermodynamic 
limit will be indicated with $f=f(\beta)$
\begin{eqnarray}
\label{f}
f(\beta)\equiv \lim_{N\rightarrow \infty}F(\beta)/N.
\end{eqnarray} 

\section{Infinite dimensionality}
We recall that a path, finite or infinite, is defined as a sequence,
finite or infinite, of 
connected bonds $\{b_{n}\}$ with no vertex repetition.
Given a set of links $\Gamma$, let us consider
the set of all the possible paths of length $l$ over $\Gamma$,
whose cardinality will be indicated with $c_N(l)$.
For random systems, important information are
contained in the probability ${p}(l_1,l_2;l)$
that two randomly chosen paths of given lengths $l_1$ and $l_2$, overlap
each other for $l\leq\mathrm{min}\{l_1,l_2\}$ bonds.
In the finite system of size $N$ we have
\begin{eqnarray}
\label{PROB}
{p}_N(l_1,l_2;l)=
\frac{c_N(l_1,l_2;l)}{c_N(l_1,l_2)},
\end{eqnarray} 
where $c_N(l_1,l_2)=c_N(l_1)c_N(l_2)$ and $c_N(l_1,l_2;l)$ 
represent the number of couples of paths of length $l_1$ and $l_2$
and the number of couples of paths of length $l_1$ and $l_2$
which overlap for $l$ bonds, respectively.  
From now on we will assume that the following limit exists
\begin{eqnarray}
\label{PROB1}
{p}(l_1,l_2;l)\equiv
\lim_{N\to \infty}{p}_N(l_1,l_2;l).
\end{eqnarray} 
The existence of this limit is a natural requirement
which has to be satisfied in order to have a thermodynamic limit
and it is in fact satisfied as soon as the vertices of
the graph $\Gamma$ become statistically equivalent for $N\to\infty$.
A quite different question concerns instead the existence
of the limits of ${p}(l_1,l_2;l)$ with respect to
$l_1$ and $l_2$. In fact, as we shall show later, at least
in finite $D$-dimensional hypercube lattice, such limits
do not exist. 

\begin{figure}[t]
\centering
\includegraphics[width=0.4\columnwidth,clip]{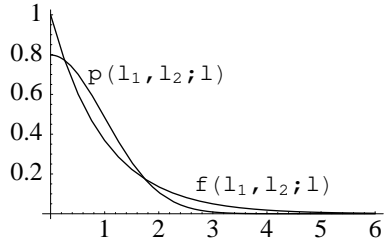}
\caption{Example of a distribution $p(l_1,l_2;l)$ which,
for $l\geq l_0=2$, is bounded from above
by an exponential distribution $f(l_1,l_2;l)$ with exponent $a=1$ 
independent from $l_1,l_2$.}
\label{probfig}
\end{figure}

We will say that the graph $\Gamma$ is \textit{infinite dimensional
in broad sense} if there exists a constant $a>0$ and $\forall l_1,l_2$
there exist two positive values $l^{(0)}_{l_1,l_2}$ and $C_{l_1,l_2}$ such that 
\small
\begin{eqnarray}\fl
\label{PROB2}
{p}(l_1,l_2;l)\leq C_{l_1,l_2} e^{-al}, 
\quad \forall l>l^{(0)}_{l_1,l_2}, 
\quad C_{l_1,l_2} \sum_{l'=0}^{\mathrm{\mathrm{min}\{l_1,l_2\}}}e^{-al'}=1,
\quad \lim_{l_1,l_2\to\infty}l^{(0)}_{l_1,l_2}<\infty.
\end{eqnarray} 
\normalsize
The above condition expresses the fact that the probability
${p}(l_1,l_2;l)$ is above bounded by an exponential
distribution, see Fig. \ref{probfig}.
Notice that $C_{l_1,l_2}\to (e^b-1)/e^b$ for $l_1,l_2\to\infty$,
so that the exponential distribution is asymptotically independent
from $l_1,l_2$.
Note also that, in general, Eq. (\ref{PROB2}) 
does not imply the existence of the limit
\begin{eqnarray}
\label{PROB2l}
p(l)\equiv \lim_{l_1,l_2 \to \infty}~
\lim_{N\to \infty}{p}_N(l_1,l_2;l) =
\lim_{l_1,l_2 \to \infty}{p}(l_1,l_2;l),
\end{eqnarray}
furthermore, even if the limit (\ref{PROB2l}) does exist,
in general it is not a probability. However, as we shall show later, in 
many important cases of interest the limit $p(l)$ exists and it is a probability.
Notice that the order of the limits in Eq. (\ref{PROB2l}) 
cannot be exchanged. 
On the other hand, it should be clear that this limit represents
the only sensible limit, being obviously $l_1,l_2\leq N$.

In the literature another definition of dimensionality is often
given. Chosen an arbitrary vertex as reference, \textit{the root vertex}, which
for $N$ large is supposed to be statistically equivalent to any
other vertex,
let $\mathcal{N}_N(l)$ be the total number of vertices at distance $l$ from
the root vertex, where the distance between two vertices 
is defined by the shortest path connecting the two vertices.
Then, by looking at the hypercube $D$-dimensional lattice of size $l$
whose dimension satisfies the rule $N\propto l^D$, by analogy,
the natural dimension of the graph $D(l)$ can be defined as
\begin{eqnarray}
\label{PROB4}
D_{\mathrm{v}}(l)\equiv 
\lim_{N\to\infty}\frac{\log\left(\mathcal{N}_N(l)\right)}{\log(l)}.
\end{eqnarray} 
It is easy to see for example \cite{Baxter} that in a Bethe lattice 
of degree $k$ one has 
$D_{\mathrm{v}}(l)=l\log(k-1)/\log(l)$, so that
in particular $D_{\mathrm{v}}(l)\to\infty$ for $\l\to\infty$.
Despite this definition of dimensionality has an intuitive
meaning, it seems not useful for practical calculations in statistical mechanics.
In fact, its use, to the best of our knowledge, has remained only
at an heuristic level, whereas, as we shall show, 
the definition of infinite dimensionality in the broad sense
we have above introduced turns out to be the sufficient
condition for the mapping to become exact which, in particular, 
allows to establish rigorously that almost all models built up over random graphs 
satisfying this condition have in fact, as
expected, a mean-field-like critical behavior \cite{MOIV}.

The definition of infinite dimensionality in the sense of Eq. (\ref{PROB4})
can be read as a definition of infinite dimensionality concerning vertices.
It is easy to check that our definition of infinite dimensionality 
implies a corresponding infinite dimensionality 
in the sense of paths as in Eq. (\ref{PROB4}) with $\mathcal{N}_N(l)$ 
replaced by $c_N(l)$. In fact, from Eq. (\ref{PROB2}) 
applied for the choice $l=l_1=l_2>l_0$ we have
\begin{eqnarray}
\label{PROB5}
c_N(l)\leq c_N(l)^2 C_{l,l} e^{-al}, \quad \forall l>l_0,
\end{eqnarray} 
where we have used the fact that $c_N(l_1,l_2)=c_N(l_1)c_N(l_2)$ and
$c_N(l,l;l)=c_N(l)$. By taking the logarithm we get
\begin{eqnarray}
\label{PROB5}
D_{\mathrm{p}}(l)\equiv \lim_{N\to\infty}
\frac{\log\left(c_N(l)\right)}{\log(l)}\geq \frac{al}{\log(l)}.
\end{eqnarray} 
Furthermore, since $c_N(l)\geq\mathcal{N}_N(l)$ for any $l$, we have also
\begin{eqnarray}
\label{PROBDIS}
D_{\mathrm{p}}(l) \geq D_{\mathrm{v}}(l)
\end{eqnarray} 
so that the infinite dimensionality in the sense of vertices 
$D_{\mathrm{v}}(l)$,
implies the infinite dimensionality in the sense of paths 
$D_{\mathrm{p}}(l)$, but not vice-versa,
the two definitions of infinite dimensionality (\ref{PROB4}) and 
(\ref{PROB5}) becoming equivalent only
in tree like structures or in structures where the number of loops
per vertex is sufficiently low (our use of the word \textit{broad} comes from this fact).
However, we stress that even if Eq. (\ref{PROB5}) is satisfied, the
condition of infinite dimensionality in the broad sense as expressed by
Eq. (\ref{PROB2}), represents a more restrictive condition
than Eq. (\ref{PROB5}),
giving a key information about the probability ${p}(l_1,l_2;l)$; in fact
Eq. (\ref{PROB5}) in general does not imply Eq. (\ref{PROB2}).

\section{Examples of infinite dimensional graphs}
\label{Examples}
Here we give a short list of examples of graphs satisfying the condition (\ref{PROB2}).
For these cases there exists even the limit distribution $p(l)$ of Eq. (\ref{PROB2l})
whose behavior can be easily estimated for large $l$. 
It is useful to keep in mind that Eq. (\ref{PROB}) lends itself to be interpreted
also as the ratio between $\tilde{c}_N(l_1,l_2;l)\equiv {c}_N(l_1,l_2;l)/N$ 
and $\tilde{c}_N(l_1,l_2)\equiv {c}_N(l_1,l_2)/N$, \textit{i.e.}, the ratio involving
the cardinalities per vertex rather than the total cardinalities.

\subsection{Regular Bethe lattice}
For a Bethe lattice of degree $k>2$ we can estimate $p(l)$ as follows.
Let us fix a root vertex and an arbitrary infinitely long path (the path 1) starting
from this root vertex. Now, let us draw out
an arbitrary infinitely long path (the path 2) starting from the root vertex. 
The path 2 can overlap the first bond of the path 1 with a probability $1/k$.
If this happens, the path 2 can overlap the path 1 over the second bond 
with a probability $1/(k-1)$ and so on. 
Therefore we have that the two paths overlap for at least l bonds with
a probability given by $p(l'\geq l)=C/(k-1)^{l-1}$, where $C$ is the normalization
constant. Since $p(l)\leq p(l'\geq l)$ and $k>2$ we have Eq. (\ref{PROB2}).
\subsection{Bethe lattices}
We can immediately generalize the above result to the case of an arbitrary infinite
tree, that is a general Bethe lattice. In this case, if the minimum statistically relevant
(that is with non zero weight) degree of the tree is $k_{\mathrm{min}}>2$, for large $l$
we have again $p(l)\leq C/(k_{\mathrm{min}}-1)^{l}$.
\subsection{Generalized tree-like structures} 
It is easy to see that even if we add a finite number of closed paths per vertex to the previous tree
(our definition of generalized tree-like structure)
for large $l$ we are again left with $p(l)\leq C/(k_{\mathrm{min}}-1)^{l}$.
\subsection{Husimi trees}
Here we note only that essentially for large $l$ we have again   
$p(l)\leq C/(k_{\mathrm{eff}}-1)^{l}$, where $k_{\mathrm{eff}}>2$.
However, we advise the reader that for such graphs the density free energy $f$ 
does not exist whereas the existence of $f$ (\textit{i.e.}, the existence of a well
defined thermodynamic) is a necessary condition for the mapping.
Therefore, in general the application of the mapping on Husimi trees may lead to wrong results.  
We will come back on this question in Sec. \ref{Nishi}. 
Nevertheless the Husimi trees provide an interesting example  
belonging to the class of non tree-like graphs mentioned in the 
introduction: these graphs are infinite dimensional in the broad sense 
but have a number of closed paths per vertex infinite. 
It would be of great interest to study a some random version of these graphs
defined in such a way that the density free energy $f$ turns out to be well defined, like
happens for the random versions of the Bethe lattices \cite{MezardP}.
We point out also that, unlike the non existence of $f$ as a proper
thermodynamic limit of a succession $f_N$ of density free energies for
finite Husimi trees of size $N$, the thermodynamic limit (\ref{PROB1}) does exist.

\subsection{Random graphs}
The previous cases can be read as examples of quenched graphs. 
Many interesting models are instead built over an ensemble of random graphs $G$
equipped with some probability $P(g)$ for extracting a given graph $g\in G$
(see for example \cite{Dorogovstev}).
In this case the free energy of the model is defined as
\begin{eqnarray}
\label{logZG}
-\beta F\equiv \sum_{g\in G}P(g)\int d\mathcal{P}\left(\{J_b\}\right)
\log\left(Z_g\left(\{J_b\}\right)\right),
\end{eqnarray} 
where $Z_g\left(\{J_b\}\right)$ 
is the partition function of the quenched system in the graph $g$
with couplings $\{J_b\}$.
Let us consider a given graph $g$ drawn out from the ensemble $G_N$ with
$N$ vertices.
At least for uncorrelated random graphs (graphs where there are no degree-degree correlations), 
it has been proved \cite{S,S1} that, if $N^s$ represents the size of the non tree-like portion 
of $g$, one has $\lim_{N\to\infty} N^s/N=0$ with probability 1.
Since the condition (\ref{PROB2}) concerns the behavior for large $l$ 
of the limit $\lim_{N\to\infty}\lim_{l_1,l_2\to\infty}p_N(l_1,l_2;l)$, 
also for the uncorrelated random graphs we can evaluate $p(l)$ 
as in the tree-like case and - again -
an exponential decay for $p(l)$ is found as soon as the mean connectivity $\bar{k}$
is greater than 2. 

\section{The mapping}
Given a spin glass model trough Eqs. (\ref{Gamma1}-\ref{dP}), 
we define, on the same set of links $\Gamma$, its \textit{related Ising model} 
trough the following Ising Hamiltonian
\begin{eqnarray}
\label{HI}
H_I\left(\{\sigma_i\};\{J_b\};\{h_i\}\right)
\equiv -\sum_{b\in\Gamma} J_b^{(I)} \tilde{\sigma}_b+\sum_{i=1}^N h_i \sigma_i
\end{eqnarray} 
where the Ising couplings $J_b^{(I)}$'s have 
non random values such that $~\forall ~b,b'\in \Gamma$
\begin{eqnarray}
\label{JI}
J_{b'}^{(I)}&=&J_b^{(I)} \quad \mathrm{if} \quad 
d\mu_{b'}\equiv d\mu_{b}, \\
\label{JIb}
J_b^{(I)}&\neq & 0 \quad \mathrm{if} \int d\mu_b(J_b)J_b\neq 0 \quad
\mathrm{or} \quad \int d\mu_b(J_b)J_b^2>0. 
\end{eqnarray}
In the following a suffix $I$ over quantities such as $H_{I}$,
$F_{I}$, $f_{I}$, etc\ldots, or $J_b^{(I)}$, $\beta_c^{(I)}$, etc\ldots,
will be referred to the related Ising system with Hamiltonian (\ref{HI}).

Let us suppose that the graph $\Gamma$ is infinite dimensional in the broad sense 
\small
\begin{eqnarray}
\label{PROB2mapp}\fl
{p}(l_1,l_2;l)\leq C_{l_1,l_2} e^{-al}, 
\quad \forall l>l^{(0)}_{l_1,l_2}, 
\quad C_{l_1,l_2} \sum_{l'=0}^{\mathrm{\mathrm{min}\{l_1,l_2\}}}e^{-al'}=1,
\quad \lim_{l_1,l_2\to\infty}l^{(0)}_{l_1,l_2}<\infty,
\end{eqnarray} 
\normalsize
and that there exists $f_I$, the density free energy of the related Ising
model in the thermodynamic limit $f_I=\lim_{N\to\infty} f_{I,N}$.
Let be $z_b^{(I)}=\tanh(\beta J_b^{(I)})$ and let 
\begin{eqnarray}
\label{GI}
G(\{z_b^{(I)}\})=0
\end{eqnarray}
represents the equation (possibly vectorial) for the critical surface 
of the related Ising model. 
Equation (\ref{GI}) describes a transition between the P phase and
an ordered F/AF phase. In the following we will 
show that $\beta_c$, the inverse of the critical temperature 
of the upper critical surface of the spin glass model is given by
\begin{eqnarray}
\label{mappg}
\beta_c=\mathrm{min}\{\beta_c^{\mathrm{(SG)}},\beta_c^{\mathrm{(F/AF)}}\}
\end{eqnarray} 
where $\beta_c^{\mathrm{(SG)}}$ and $\beta_c^{\mathrm{(F/AF)}}$ are the solutions of the two following
equations 
\begin{eqnarray}
\label{mapp0g}
G_I\left(\left\{\media{\tanh^2(\beta_c^{\mathrm{(SG)}} J_b)}\right\}\right)&=& 0,\\
\label{mapp01g}
G_I\left(\left\{\media{\tanh(\beta_c^{\mathrm{(F/AF)}} J_b)}\right\}\right)&=& 0.
\end{eqnarray} 

Note that Eqs. (\ref{mappg}) - (\ref{mapp01g}) describe completely 
the upper critical surface. 
So for example, in a case with two families of bonds $b_1$
and $b_2$, whose couplings $J_1$ and $J_2$ are 
distributed according to the measures $d\mu_1$ and $d\mu_2$, respectively,
the equation 
$G(\int d\mu_1\tanh(\beta J_1),\int d\mu_2\tanh^2(\beta J_2))=0$ 
does not describe any upper
critical surface; for the upper critical surface
there are no intermediate situations between 
Eqs. (\ref{mapp0g}) and (\ref{mapp01g}).

Equations (\ref{mappg}) - (\ref{mapp01g}) give the exact 
critical P-SG and P-F/AF temperatures.
In the case of a measure $d\mu$ independent on the bond $b$, the suffix F or AF stands
for disordered ferromagnetic or antiferromagnetic phase, respectively.
In the general case, such a distinction is possible only
in the positive and negative sectors where one has respectively 
$\media{\tanh(\beta_c^{\mathrm{(F/AF)}} J_b)}>0$ or 
$\media{\tanh(\beta_c^{\mathrm{(F/AF)}} J_b)}<0$, for any bond $b$, whereas, for
the other sectors, we use the symbol
F/AF only to stress that the transition is not P-SG.

Near the upper critical surface, at zero external field,
the mapping allows also to determine the correlation functions exactly.
We remind the reader to the Ref. \cite{MOI} for details (we stress here that
Eqs. (38-41) of Ref. \cite{MOI} at zero external field are exact).

\section{High temperature expansion}
Let us consider a generic Ising model 
at zero external field with given couplings $\{J_b\}$ 
defined over some set of links $\Gamma$. Note that, since the couplings are
arbitrary, what we will say will be valid, in particular, for the
related Ising model.  
It is convenient to introduce the symbol
\begin{eqnarray}
K_b\equiv \beta J_b.
\end{eqnarray} 
For the partition function it holds the so called 
``high temperature'' expansion
\begin{eqnarray}
\label{Z2}
Z\left(\{J_b\}\right)= \prod_{b\in\Gamma} 
\cosh\left(K_b\right) \sum_{\{\sigma_i\}}
\prod_{b\in\Gamma} \left(1+\tilde{\sigma}_b\tanh\left(K_b\right)\right). 
\end{eqnarray} 
As is known the terms obtained by expansion of the product 
$\prod_{b\in\Gamma} \left(1+\tilde{\sigma}_b\tanh\left(K_b\right)\right)$,
with $k$ bonds proportional to 
$\tilde{\sigma}_{b_1}\tilde{\sigma}_{b_2}\ldots \tilde{\sigma}_{b_k}$, 
contribute to the sum over the spins only
if the set $\gamma\equiv \{b_1,b_2,\ldots,b_k\}$ constitutes 
a closed multi-polygon over $\Gamma$ for open or periodic boundary 
conditions,
and a collection of multi-polygons and paths, 
whose end-points belong to the boundary of $\Gamma$,
for closed conditions (when all the spins on the boundary are fixed 
to be +1 or -1) (\textit{e.g.}, see \cite{GG,Domb}); 
in such cases $\tilde{\sigma}_{b_1}\tilde{\sigma}_{b_2}\ldots 
\tilde{\sigma}_{b_k}\equiv 1$
so that Eq. (\ref{Z2}) becomes
\begin{eqnarray}
\label{Z3}
Z\left(\{J_b\}\right)= 2^{N} \prod_{b\in\Gamma} 
\cosh\left(K_b\right) \sum_{\gamma} 
\prod_{b\in \gamma}\tanh\left(K_b\right), 
\end{eqnarray} 
where the sum runs over all the above mentioned multi components paths,
shortly multi-paths, $\gamma$.
Note that in the case $\tanh\left(K_b\right)=0$, 
the sum over the paths gives 1, (\textit{i.e.} the contribution
with zero paths must be included). 

From Eq. (\ref{Z3}) we have
\begin{eqnarray}\fl
\label{logZ1}
\int d\mathcal{P}\left(\{J_b\}\right)\log\left(Z\left(\{J_b\}\right)\right)&=&
\int d\mathcal{P}\left(\{J_b\}\right)
\log\left(2^N\prod_{b\in\Gamma} \cosh(K_b)\right) \nonumber \\ && 
+ \int d\mathcal{P}\left(\{J_b\}\right)
\log\left(\sum_{\gamma} \prod_{b\in \gamma}
\tanh(K_{b})\right),
\end{eqnarray} 
from which, by using Eqs. (\ref{logZ}) and (\ref{dP}) 
in the first term of the r.h.s., we get
\begin{eqnarray}
\label{logZ2}
-\beta F=N\log(2)+
\sum_{b\in\Gamma} \int d\mu_{b} \log\left(\cosh(K_b)\right) + \phi,
\end{eqnarray}
where the non trivial part $\phi$ is given by
\begin{eqnarray}
\label{phi0}
\phi \equiv \int d\mathcal{P}\left(\{J_b\}\right)\log\left(\sum_{\gamma} 
\prod_{b\in \gamma}\tanh(K_{b})\right).
\end{eqnarray} 
With the symbol $\phi_I\left(\{z_b^{(I)}\}\right)$ we will mean the
non trivial part of the free energy of the related Ising model 
\begin{eqnarray}
\label{phiI}
\phi_I \left(\{z_b^{(I)}\}\right)\equiv
\log\left(\sum_{\gamma} 
\prod_{b\in \gamma}z_b^{(I)}\right).
\end{eqnarray} 
The densities of $\phi$ and $\phi_I$ will be indicated as $\varphi$ 
and $\varphi_I$, respectively:
\begin{eqnarray}
\label{varphi}
\varphi \equiv \lim_{N\to\infty}\frac{\phi}{N}.
\end{eqnarray} 
\begin{eqnarray}
\label{varphiI}
\varphi_I \equiv \lim_{N\to\infty}\frac{\phi_I}{N},
\end{eqnarray} 

There exist series expansion over suitable graphs also
for $\varphi_I$ or $\varphi$, see \cite{Domb}. 
We will suppose $\varphi_I$ to be known and we will derive
$\varphi$ in terms of $\varphi_I$.
Note that, unlike the series for $\varphi_I$ and $\varphi$,
in the thermodynamic limit the two series 
\begin{eqnarray}
\label{P}
P_{}\left(\{z_b\}\right)\equiv \sum_{\gamma} \prod_{b\in \gamma} z_b,
\end{eqnarray} 
and
\begin{eqnarray}
\label{PI}
P_{}\left(\{z_b^{(I)}\}\right)\equiv 
\sum_{\gamma} \prod_{b\in \gamma} z_b^{(I)},
\end{eqnarray} 
\textit{i.e.}, the series inside the logarithm 
of the r.h.s. of Eqs. (\ref{phi0}) and
(\ref{phiI}), respectively, diverge. 
However, for establishing the mapping 
we find much more convenient to work directly with the 
series $P\left(\{z_b\}\right)$ and $P\left(\{z_b^{(I)}\}\right)$ 
to be thought as formal series. 
Given two series of the kind (\ref{P}) or (\ref{PI}), 
to our aims it will be sufficient to show that the two series coincide 
term by term.
The only important thing to note here is that
the series for $\varphi$ ($\varphi_I$) 
will be convergent for values 
of the parameters $z_b=\tanh(\beta J_b)$ ($z_b^{(I)}=\tanh(\beta J_b^{(I)})$)
sufficiently small,
\textit{i.e.}, inside a suitable set $\mathscr{D}$ ($\mathscr{D}_I$)
whose boundary corresponds to a critical surface $\Sigma$ ($\Sigma_I$) 
of the model. Since we have 
power series with positive coefficients, it turns out
that $\mathscr{D}$ ($\mathscr{D}_I$), is a convex set.
Furthermore, as already explained in the Ref. \cite{MOI}
it is clear that
\begin{center}
\textit{The critical behavior of the system is determined by
the paths of arbitrarily large length}
\end{center}

\section{Averaging over the disorder}
Let us now average $P_{}$ over the quenched couplings (the disorder)  
\begin{eqnarray}
\label{PA}
P^{(1)}\left(\{F_b^{(1)}\}\right) \equiv \int d\mathcal{P}\left(\{J_b\}\right) 
P_{}\left(\{\tanh(K_b)\}\right), 
\end{eqnarray} 
where we have introduced
\begin{eqnarray}
F_b^{(1)} \equiv \int d\mu_b \tanh(K_b).
\end{eqnarray} 
From the product nature of the distribution 
$d\mathcal{P}\left(\{J_b\}\right)$, 
Eq. (\ref{dP}), it is immediate
to see that $P^{(1)}$ is given in terms of the  
function $P_{}$ through 
\begin{eqnarray}
\label{PA1}
P^{(1)}\left(\{F_b^{(1)}\}\right)=P_{}\left(\{F_b^{(1)}\}\right)=
\sum_{\gamma} \prod_{b\in \gamma} F_b^{(1)}.
\end{eqnarray} 

Later, to evaluate the free energy 
we will need to consider also the averages 
of $P_{I}^n$ for $n\in \mathop{\rm N}$
\begin{eqnarray}
\label{PAn}
P^{(n)}\left(\{F_b^{(1)},\ldots,F_b^{(n)}\}\right) \equiv 
\int d\mathcal{P}\left(\{J_b\}\right) P_{}^{n}\left(\{\tanh(K_b)\}\right), 
\end{eqnarray} 
where for $m=1,\ldots,n$ we have introduced
\begin{eqnarray}
\label{Fn}
F_b^{(m)} \equiv \int d\mu_b \left(\tanh(K_b)\right)^m.  
\end{eqnarray} 
Note that, according to Eqs. (\ref{JI}-\ref{JIb}),
the function $P_{}\left(\{F_b^{(m)}\}\right)$ is the non trivial
part of the high temperature expansion of the related Ising model
with couplings $\{F_b^{(m)}\}$.

Let us now generalize Eq. (\ref{PA1}) to $P^{\left(n\right)}$. 
From Eqs. (\ref{P}) and (\ref{PAn}) we see that for
$n$ integer we can calculate  $P^{\left(n\right)}$ by summing over $n$ replicas
of paths $\gamma_1,\ldots,\gamma_n$, specifying for any of their
bonds how many overlaps are there with all the other paths. 
We arrive then at the following expression (see Fig. 1) 
\begin{eqnarray}
\label{PAn1}
P^{\left(n\right)}&=&  
\sum_{\gamma_1,\ldots,\gamma_n} \int d\mathcal{P}\left(\{J_b\}\right) 
\prod_{~~b \in \cap_{l=1}^{n} \gamma_{l}} \tanh^{n}\left(K_b\right) 
\times \nonumber \\
&& \prod_{(i_1)} \prod_{~~b \in \cap_{l=1,l\neq i_1}^{n} \gamma_{l} 
\setminus \gamma_{i_1}} 
\tanh^{\left(n-1\right)}\left(K_b\right) \times \nonumber \\
&& \prod_{\left(i_1,i_2\right)} \prod_{~~b \in \cap_{l=1,l\neq i_1,i_2}^{n} 
\gamma_{l}
\setminus \left(\gamma_{i_1} \cup \gamma_{i_2}\right)} 
\tanh^{\left(n-2\right)}\left(K_b\right) \dots \times \nonumber \\
&& \prod_{(i_n)} \prod_{~~b \in \gamma_{i_n} 
\setminus \left( \cup_{l \neq i_n} \gamma_l \right) } \tanh\left(K_b\right),
\end{eqnarray} 
where, in the product $\prod_{\left(i_1,i_2,\ldots i_k\right)}$, the 
indices $i_1,i_2,\ldots,i_k$ run 
over the $n!/\left(\left(n-k\right)!k!\right)$ combinations to arrange 
$k$ numbers from the integers $1,\ldots,n$, and with the symbol 
$(i_1,i_2)$ we mean the
couple $i_1,i_2$ with $i_1\neq i_2$ and similarly for 
$\left(i_1,i_2,\ldots i_k\right)$.
From Eq. (\ref{PAn1}) by using Eq. (\ref{dP}) 
and the definitions (\ref{Fn}), we arrive at 
\begin{eqnarray}
\label{PAn2}
P^{\left(n\right)}&=& \sum_{\gamma_1,\ldots,\gamma_n} 
\prod_{~~b \in \cap_{l=1}^{n} \gamma_{l}} F^{\left(n\right)}_{b} 
\times \nonumber \\
&& \prod_{(i_1)} \prod_{~~b \in \cap_{l=1,l\neq i_1}^{n} \gamma_{l} 
\setminus \gamma_{i_1}} F^{\left(n-1\right)}_{b} \times \nonumber \\
&& \prod_{\left(i_1,i_2\right)} \prod_{~~b \in \cap_{l=1,l\neq i_1,i_2}^{n} 
\gamma_{l}
\setminus \left(\gamma_{i_1} \cup \gamma_{i_2}\right)} 
F^{\left(n-2\right)}_{b} \times \dots \times \nonumber \\
&& \prod_{(i_n)} \prod_{~~b \in \gamma_{i_n} 
\setminus \left( \cup_{l \neq i_n} \gamma_l \right) } F^{\left(1\right)}_{b}.
\end{eqnarray} 

The free energy density term $\varphi$ will be obtained in terms of $P^{(n)}$, 
Eq. (\ref{PAn}), via the replica method by using the relation
\begin{eqnarray}
\label{phi1}
\varphi = \lim_{n \rightarrow 0}\lim_{N \rightarrow \infty} 
\frac{P^{(n)}-1}{nN}.
\end{eqnarray} 
Note that, as usually done in the context of a replica
approach, we have assumed that the limit $n\to 0$ and $N\to \infty$
may be exchanged (in \cite{Hemmen} it has been proved for
the Sherrington-Kirkpatrick model).

\section{Proof of the mapping}
%
Let us now consider a finite system with $N$ spins (from now on, we will
add a suffix $N$ to indicate this).
The proof we will give it remains true for any distribution $d\mu_b$, 
however, for the sake of simplicity,
we will see the proof in detail for the case of a 
homogeneous measure $d\mu$ (same measure for any bond).

Given $n$ arbitrary multi component paths, 
shortly $n$ multi-paths, or $n$ replica multi-paths,
we will say that a bond $b$ forms an $m$-overlap among the $n$ multi-paths, 
with $m\leq n$,
if it belongs exactly to $m$ of the $n$ multi-paths and we will
indicate with $l^{(m)}$ the total number of bonds forming an $m$-overlap among
the $n$ multi-paths (see Fig. 1). 
Let us consider the set of all the possible $n$ multi-paths 
with fixed length $l_1,\ldots,l_{n}$ 
whose cardinality will be indicated by $C_N(l_1,\ldots,l_{n})$
and let us consider the subset of all the possible $n$ multi-paths 
having $l^{(2)}$ 2-overlaps,\ldots ,$l^{(m)}$ $m$-overlaps, $m\leq n$, and let
$C_N(l_1,\ldots,l_{n};l^{(2)},\ldots,l^{(m)})$ be its cardinality.

Clearly, the following quantity
\begin{eqnarray}
\label{PROB6}
\mathcal{P}_N(l_1,\ldots,l_{n};l^{(2)},\ldots,l^{(k)})\equiv
\frac{C_N(l_1,\ldots,l_{n};l^{(2)},\ldots,l^{(k)})}{C_N(l_1,\ldots,l_{n})},
\end{eqnarray} 
represents the probability that choosing randomly $n$ multi-paths with
length $l_1,\ldots,l_{n}$, they form
$l^{(2)}$ 2-overlaps,\ldots ,$l^{(k)}$ $k$-th overlaps.
Let us now assume that Eq. (\ref{PROB2}) is satisfied.
Note that $c_N(l_1,l_2)$, $c_N(l_1,l_2;l^{(2)})$ and ${p}_N(l_1,l_2;l^{(2)})$ 
in Eqs. (\ref{PROB}-\ref{PROB2}) refer to 
paths, or more precisely, to simple connected paths, whereas 
$C_N(l_1,l_2)$, $C_N(l_1,l_2;l^{(2)})$, $\mathcal{P}_N(l_1,l_2;l^{(2)})$
in Eqs. (\ref{PROB6}) refer to multi-paths so that, in general, 
$\mathcal{P}_N(l_1,l_2;l^{(2)})\neq {p}_N(l_1,l_2;l^{(2)})$.
However, in Appendix A we show that if Eq. (\ref{PROB2}) is satisfied, 
for $l$ large enough we have also
\begin{eqnarray}
\label{PROB7}
\mathcal{P}(l_1,l_2;l)\leq C_{l_1,l_2}'e^{-a'l}, \quad \forall l_1,l_2, 
\end{eqnarray} 
where $a'$ is a positive constant and $C_{l_1,l_2}'$ is the normalization
(becoming also a constant in the limit $l_1,l_2\to\infty$).

Let us consider the terms with even index 
$P^{(0)}_N,P^{(2)}_N,P^{(4)}_N,\ldots,P^{(2n)}_N$.
Using the fact that the measure is the same for any bond, we can
rewrite Eqs. (\ref{PAn2}) in terms of the $C_N$'s as follows 
\begin{eqnarray}
\label{PEVEN0}\fl
P^{(0)}_N &=& 1,
\end{eqnarray} 
\begin{eqnarray}
\label{PEVEN20}\fl
P^{(2)}_N &=& \sum_{l_1,l_2;l^{(2)}}
C_N(l_1,l_2;l^{(2)}) 
\left(F^{\left(1\right)}_{b}\right)^{l^{(1)}}\left(F^{\left(2\right)}_{b}\right)^{l^{(2)}}
\end{eqnarray} 
\begin{eqnarray}
\label{PEVEN40}\fl
P^{(4)}_N &=& \sum_{l_1,l_2,l_3,l_4;l^{(2)},l^{(3)},l^{(4)}}
C_N(l_1,l_2,l_3,l_4;l^{(2)},l^{(3)},l^{(4)}) 
\left(F^{\left(1\right)}_{b}\right)^{l^{(1)}}\nonumber \\
&& \times \left(F^{\left(2\right)}_{b}\right)^{l^{(2)}}
\left(F^{\left(3\right)}_{b}\right)^{l^{(3)}}
\left(F^{\left(4\right)}_{b}\right)^{l^{(4)}},
\end{eqnarray} 
\begin{eqnarray}
\label{PEVEN2n0}\fl
P^{(2n)}_N &=& \sum_{l_1,\ldots,l_{2n};l^{(2)},\ldots,l^{(2n)}}
C_N(l_1,\ldots,l_{2n};l^{(2)},\ldots,l^{(2n)}) 
\left(F^{\left(1\right)}_{b}\right)^{l^{(1)}}\nonumber \\
&& \times \ldots \left(F^{\left(2n\right)}_{b}\right)^{l^{(2n)}},
\end{eqnarray} 
where we have made use of the shorter notation $l^{(1)}$ for the bonds
with no overlap, determined by the other lengths as 
\begin{eqnarray}
\label{l1}
l^{(1)}\equiv l_1+\ldots+l_{2n}-2l^{(2)}-\ldots 2nl^{(2n)}. 
\end{eqnarray} 

\subsection{Symmetric measure}
Let us consider a symmetric measure:
\begin{eqnarray}
\label{dmusim}
d\mu_b(-J_b)=d\mu_b(J_b).
\end{eqnarray}   
The symmetric measure represents the most difficult case.
Once analyzed this case the problem for a general measure
will be easily derived.
\subsubsection{First step}\label{First}
For a symmetric measure we have
\begin{eqnarray}
\label{Fsimm}
F^{\left(2m+1\right)}_{b}= 0.
\end{eqnarray} 
As a consequence, we see that the only non zero contributions
are those having zero odd-overlaps: $l^{(1)}=l^{(3)}=\ldots=l^{(2n-1)}=0$,
so that Eqs. (\ref{PEVEN20}-\ref{PEVEN2n0}) become 
\begin{eqnarray}
\label{PEVEN201}\fl
P^{(2)}_N &=& \sum_{l}
C_N(l) \left(F^{\left(2\right)}_{b}\right)^{l}=P_N\left(F^{\left(2\right)}_{b}\right),
\end{eqnarray} 
\begin{eqnarray}
\label{PEVEN401}\fl
P^{(4)}_N &=& \sum_{l_1,l_2,l_3,l_4;l^{(2)},l^{(4)}}
C_N(l_1,l_2,l_3,l_4;l^{(2)},0,l^{(4)})\delta_{l^{(1)},0} 
\left(F^{\left(2\right)}_{b}\right)^{l^{(2)}}
\left(F^{\left(4\right)}_{b}\right)^{l^{(4)}},
\end{eqnarray} 
\begin{eqnarray}
\label{PEVEN2n01}\fl
P^{(2n)}_N &=& \sum_{l_1,\ldots,l_{2n};l^{(2)},l^{(4)},\ldots,l^{(2n)}}
C_N(l_1,\ldots,l_{2n};l^{(2)},0,l^{(4)},\ldots,0,l^{(2n)}) 
\nonumber \\
&& \times \delta_{l^{(1)},0} 
\left(F^{\left(2\right)}_{b}\right)^{l^{(2)}}\left(F^{\left(4\right)}_{b}\right)^{l^{(4)}}
\ldots 
\left(F^{\left(2n\right)}_{b}\right)^{l^{(2n)}}.
\end{eqnarray} 
In deriving Eq. (\ref{PEVEN201}) we have observed that the only
non zero contributions in Eq. (\ref{PEVEN20}) are those having 
$l^{(2)}=l_1=l_2$, \textit{i.e.}, the only couples of multi-paths 
contributing to Eq. (\ref{PEVEN20}) are those which overlap completely 
two to two. Let us see now the term $P^{(4)}_N$. The constrain
$l^{(1)}=0$ implies that the 2-overlap is determined by the length of the
four multi-paths and by $l^{(4)}$ as 
\begin{eqnarray}
\label{PEVEN4l2}\fl
l^{(2)}=l^{(2)}(l_1,l_2,l_3,l_4;l^{(4)})=\frac{l_1+l_2+l_3+l_4}{2}-2l^{(4)}.
\end{eqnarray} 
The above expression for $l^{(2)}$ is made clear by Figs. 
(\ref{R}-\ref{2_and_2})
where we show all the possible topologically equivalent situations.
\begin{figure}[t]
\centering
\includegraphics[width=0.4\columnwidth,clip]{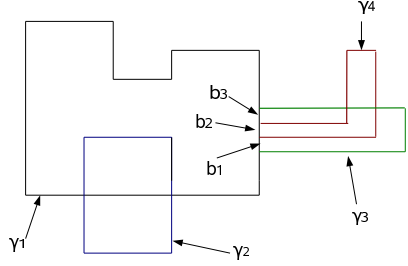}
\caption{A contribution to the summation of Eq. (\ref{PAn1}) with $n=4$.
Here we have: a bond with overlap of order 3, 
$b_2=\gamma_1\cap\gamma_3\cap\gamma_4$;
two bonds with overlap of order 2,
$b_1\cup b_3=\gamma_1\cap\gamma_3$; 
and all the other bonds with no overlap (order 1).
Note that the multi-paths $\gamma_1$ and $\gamma_2$
intersect each other as geometrical objects
but not as sets (for definition, a path $\gamma$ is the union of its bonds).
The same observation holds for the multi-paths 
$\gamma_3\setminus (b_1\cup b_2\cup b_3)$
and $\gamma_4\setminus b_2$.}
\label{4_paths}
\end{figure}
\begin{figure}[t]
\centering
\includegraphics[width=0.2\columnwidth,clip]{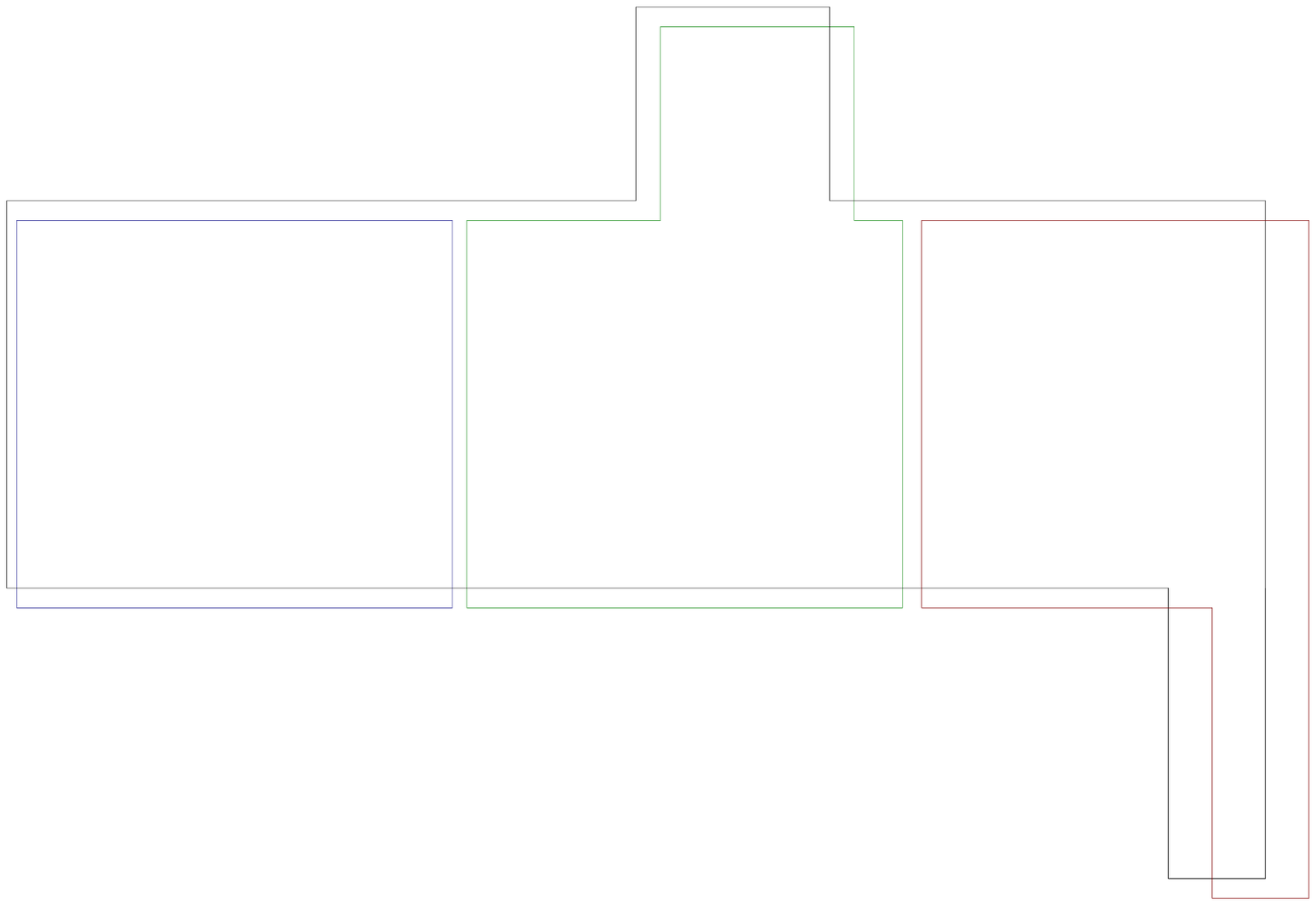}
\caption{ 
A chain of three connected planar multi-paths encapsulated in a larger planar path. 
The four multi-paths 
overlap each other only partially.
The multi-paths in the figure are slightly shifted for visual convenience.}
\label{R}
\end{figure}
\begin{figure}[t]
\centering
\includegraphics[width=0.2\columnwidth,clip]{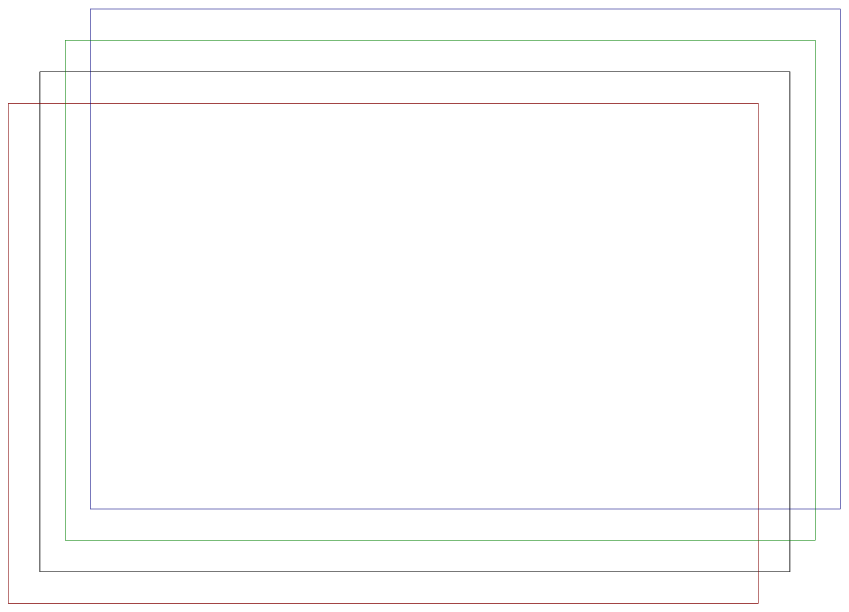}
\caption{ 
Schematic example of 
four completely overlapping planar multi-paths. 
The multi-paths in the figure are slightly shifted for visual convenience.}
\label{4_overlaps}
\end{figure}
\begin{figure}[t]
\centering
\includegraphics[width=0.2\columnwidth,clip]{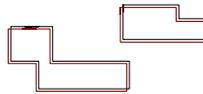}
\caption{ 
Schematic example of 
two couples of two completely overlapping planar multi-paths.
The multi-paths in the figure are slightly shifted for visual convenience.}
\label{2_and_2}
\end{figure}
We find it useful to decompose the set of the four replica multi-paths 
of length $l'\leq l$ as follows
\begin{eqnarray}
\label{E4}
\mathscr{E}_{4}(l)=\mathscr{E}_{4,4}(l)\cup\mathscr{E}_{2,4}(l)\cup\mathscr{R}_{4}(l),
\end{eqnarray}  
where $\mathscr{E}_{4,4}(l)$ is the set of 4 coinciding multi-paths, \textit{i.e.}
the set of all one replica multi-paths, 
see Fig. \ref{4_overlaps};
$\mathscr{E}_{2,4}(l)$ is the set of all multi-paths
overlapping in couples (two to two), see Fig. \ref{2_and_2};
and $\mathscr{R}_{4}(l)$ is the rest, \textit{i.e.}
the set of all multi-paths in $\mathscr{E}_{4}(l)$
which overlap each other only partially, see Fig. \ref{R}.

In \cite{MOI} we have shown that, when $D$ is large enough, in a $D$-dimensional hypercube lattice for any $l$ one has
\begin{eqnarray}
\label{E4e}
\frac{|\mathscr{E}_{4,4}(l)|}
{|\mathscr{E}_{2,4}(l)|}=
O\left(\frac{1}{D-1}\right), \quad
\frac{|\mathscr{R}_{4}(l)|}
{|\mathscr{E}_{2,4}(l)|}=
O\left(\frac{1}{D-1}\right).
\end{eqnarray}  

In a $D$-dimensional hypercube lattice any vertex is crossed by
$D$ axis, which can be seen as infinitely long non overlapping
multi-paths. Equation (\ref{E4e}) says that, for any $l$, when $D\to\infty$
the sets $\mathscr{R}_{4}(l)$ and $\mathscr{E}_{4,4}(l)$ can be completely neglected
and the mapping becomes exact.
The fact that Eq. (\ref{E4e}) holds for any $l$
allowed to consider the mapping for all temperatures where the 
high temperature expansion converges, that is for all $\beta<\beta_c$.
We want now to try to generalize Eq. (\ref{E4e}) to arbitrary graphs.
The idea is to observe that, even if the mean
connectivity $\bar{k}$ is finite, as $l$ goes to infinity,
Eq. (\ref{E4e}) becomes true. However, as we shall show soon, for general
graphs we cannot isolate the sets $\mathscr{E}_{2,4}$ and $\mathscr{E}_{4,4}$
because their union constitute the elements over which the summation runs.
%
Let us observe that, if $\bar{k}$, the mean connectivity of a vertex
is greater than 1, $\bar{k}>1$, we have that $C_N(l)$ grows exponentially in $l$. For the
total number of simple (non multi) paths one has $c_N(l)\propto (\bar{k}-1)^l$ \cite{Fisher} but
approximately this remains true also for the total number of multi-paths 
$C_N(l)=\mathrm{O}((\bar{k}-1)^l)$. However, for what follows, we do not need the exact expression for
$C_N(l)$. For our aims here, it is enough to keep in mind that $C_N(l)$ grows 
exponentially with $l$ according to some rate related to $\bar{k}$.
For any graph and for large $l$ we have 
\begin{eqnarray}
\label{NT}
|\mathcal{E}_{4,4}(l)| \propto C_N(l),
\end{eqnarray}   
where we have used the fact that
there is only one single way to
overlap completely four multi-paths of length $l$.
Similarly, for the set $\mathscr{R}_{4}(l)$ we have
\begin{eqnarray}
\label{RT}
|\mathcal{R}_4(l)|=\sum_{l'}^l\sum\limits_{l_1,l_2,l_3:~l_1+l_2+l_3=l'}
C_N(l')\propto l^3C_N(l)
\end{eqnarray}   
where we have taken into account the degeneracy coming
from the fact that given a multi-path of length $l'$,
we have to chose 3 multi-paths of lengths $l_1,l_2,l_3$ in all the possible ways 
such that be $l_1+l_2+l_3=l'$.
The set $|\mathscr{E}_{2,4}(l)|$ represents the total number of
ways to arrange two multi-paths of length $l_1\leq l$ and $l_2\leq l$
which do not share any bond each other. Hence we have
\begin{eqnarray}
\label{E2T}
{|\mathscr{E}_{2,4}(l)|}\propto \left(C_N(l)\right)^{2}-C_N(l)
\end{eqnarray}  

Therefore, from Eqs. (\ref{NT}) and (\ref{RT}) for
large $l$ we arrive at the analogous of Eqs. (\ref{E4e})
\begin{eqnarray}
\label{E4l}
\frac{|\mathscr{E}_{4,4}(l)|}
{|\mathscr{E}_{2,4}(l)|}\propto \frac{1}{C_N(l)}, \quad
\frac{|\mathscr{R}_{4}(l)|}
{|\mathscr{E}_{2,4}(l)|}\propto \frac{l^3}{C_N(l)},
\end{eqnarray}  
implying that for sufficiently large lengths,
we can neglect the contributions coming from the sets
$\mathscr{E}_{4,4}$ and $\mathscr{R}_{4}$ so that, from
Eq. (\ref{PEVEN401}), after a change of names, we are left with 
\begin{eqnarray}
\label{PEVEN40}\fl
P^{(4)}_N &=& 
3\sum_{l_1,l_2,l^{(2)}}^N C_N(l_1,l_2;l^{(2)})
\left(F^{\left(2\right)}_{b}\right)^{l_1+l_2-2l^{(2)}}
\left(F^{\left(4\right)}_{b}\right)^{l^{(2)}},
\end{eqnarray} 
where we have taken into account that we have 3 ways to couple 4 replicas
of multi-paths overlapping two to two. Note that, as anticipated, we can neglect the
contributions coming from the sets $\mathscr{E}_{4,4}$ and $\mathscr{R}_{4}$,
but, in fact, , unlike a graph
having $\bar{k}\to\infty$, as manifested by Eq.(\ref{PEVEN40}), we cannot isolate 
the sets $\mathscr{E}_{4,4}$ and $\mathscr{E}_{2,4}$. This problem will be analyzed 
in the second step of the proof.

Similarly, for $P^{(6)}$, 
one finds $|\mathscr{E}_{2,6}(l)|\propto \left(C_N(l)\right)^{3}$, 
$|\mathscr{E}_{4,6}(l)|\propto \left(C_N(l)\right)^{2}$, 
$|\mathscr{E}_{6,6}(l)|\propto C_N(l)$, 
and $|\mathscr{R}_{6}(l)|\propto l^5 \left(C_N(l)\right)^{2}$, 
so that, after a change of names,
we are left with 
\begin{eqnarray}
\label{PEVEN60}\fl
P^{(6)}_N &=& 
15\sum_{l_1,l_2,l_3,l^{(2)},l^{(3)}}^N C_N(l_1,l_2,l_3;l^{(2)},l^{(3)})
\left(F^{\left(2\right)}_{b}\right)^{l_1+l_2+l_3-2l^{(2)}-3l^{(3)}}
\nonumber \\ \fl && \times
\left(F^{\left(4\right)}_{b}\right)^{l^{(2)}}
\left(F^{\left(6\right)}_{b}\right)^{l^{(3)}},
\end{eqnarray} 
where we have taken into account that we have $15=6!/2^{3}$ 
ways to couple 6 replicas
of multi-paths two to two, but only one way to overlap the
6 replicas of multi-paths completely over the same single path. 
Generalizing to arbitrary $n$ we arrive at
\begin{eqnarray}
\label{PEVEN0}\fl
P^{(2n)}_N &=& \frac{(2n)!}{2^n}\sum_{l_1,\ldots,l_n,l^{(2)},\ldots,l^{(n)}}^N 
C_N(l_1,\ldots,l_n,l^{(2)},\ldots,l^{(n)})
\nonumber \\ \fl && \times
\left(F^{\left(2\right)}_{b}\right)^{l_1+\ldots+l_n-2l^{(2)}-\ldots-nl^{(n)}}
\left(F^{\left(4\right)}_{b}\right)^{l^{(2)}}\cdots
\left(F^{\left(2n\right)}_{b}\right)^{l^{(n)}}.
\end{eqnarray} 

If we choose the delta plus minus measure $d\mu_{\pm}$ 
\begin{eqnarray}
\label{mupm}
\frac{d\mu_{\pm}(J_b)}{dJ_b}\equiv \frac{1}{2}\delta(J_b-J)+\frac{1}{2}\delta(J_b+J),
\end{eqnarray} 
due to the property $\left(F^{\left(2n\right)}_{b}\right)=
\left(F^{\left(2\right)}_{b}\right)^n$, 
and using $\sum_{l^{(2)}}C(l_1,l_2;l^{(2)})$ we get immediately 
that near the critical temperature
\begin{eqnarray}
\label{PEVEN60Gpm}
P^{(2n)}_{N\to\infty}=
\left(P_{N\to\infty}\left(F_b^{(2)}\right)\right)^n, 
\quad d\mu(J_b)=d\mu_{\pm}(J_b).
\end{eqnarray} 

\subsubsection{Second step}\label{Second}
We want now to show that for 
any symmetrical measure $d\mu$, if Eq. (\ref{PROB2}) (and then (\ref{PROB7})) 
is satisfied, near the critical temperature it holds
\begin{eqnarray}
\label{PEVEN60G}
P^{(2n)}_{N\to\infty}\propto 
\left(P_{N\to\infty}\left(F_b^{(2)}\right)\right)^n.
\end{eqnarray} 

The idea of the proof is based on three key points.
First, we use the fact that
near the critical temperature only infinitely long multi-paths 
contribute to $P^{(n)}$.
Second, we use our fundamental hypothesis (\ref{PROB2}).
Third, we use the observation that,
given two series $S=\sum_l a_l z^l$ and $S'=\sum_l a_l b_l z^l$,
if for $l\to\infty$, $b_l\to b$ with $0<|b|<\infty$, the two series  
have the same radius of convergence. 
Let us rewrite Eq. (\ref{PEVEN40}) as
\begin{eqnarray}
\label{PEVEN4}
P^{(4)}_N &=& 3\sum_{l_1,l_2,l^{(2)}}^N C_N(l_1,l_2;l^{(2)})
\left(F^{\left(2\right)}_{b}\right)^{l_1+l_2}
x_b^{l^{(2)}},
\end{eqnarray} 
where
\begin{eqnarray}
\label{PEVEN4x}
x_b(\beta)\equiv 
\frac{F^{\left(4\right)}_{b}}{\left(F^{\left(2\right)}_{b}\right)^2}.
\end{eqnarray} 
Notice that, due to the Cauchy-Schwartz inequality, 
$x_b\geq 1$ and it is equal to 1 only for the 
delta plus-minus measure. 
To study the critical behavior of the system we need to know
the singularities of $P^{(2n)}_N$ for $N\to\infty$.
Let us now rewrite Eq. (\ref{PEVEN4}) as follows
\begin{eqnarray}
\label{PEVEN4A}
P^{(4)}_N &=& 3\sum_{l_1,l_2}^N C_N(l_1,l_2)B_N(l_1,l_2)
\left(F^{\left(2\right)}_{b}\right)^{l_1+l_2},
\end{eqnarray} 
where 
\begin{eqnarray}\fl
\label{PEVEN4B}
B_N(l_1,l_2)&\equiv&\frac{\sum_{l^{(2)}}^{\mathrm{min}(l_1,l_2)} 
C_N(l_1,l_2;l^{(2)})x_b^{l^{(2)}}}
{\sum_{l^{(2)}}^{\mathrm{min}(l_1,l_2)} C_N(l_1,l_2;l^{(2)})} 
\nonumber \\
 &=& \sum_{l^{(2)}}^{\mathrm{min}(l_1,l_2)}
\mathcal{P}_N(l_1,l_2;l^{(2)})x_b^{l^{(2)}}.
\end{eqnarray} 
and in Eq. (\ref{PEVEN4B}) we have made explicit in the sums the 
limit in $l^{(2)}$, $\mathrm{min}(l_1,l_2)$,
above which $C_N(l_1,l_2;l^{(2)})$ becomes 0.
Let us observe that in the $\lim_{l_1,l_2\to\infty}~\lim_{N\to \infty}$, 
Eq. (\ref{PEVEN4B}) becomes a power series in $l^{(2)}$ in powers of $x_b$
which, due to Eq. (\ref{PROB7}), has a non zero radius of convergence
given by $\exp(a')$.
The given measure $d\mu$ belongs to some functional space $\mathcal{L}$
embedded with some distance $\|\cdot\|$.
Let us now introduce in $\mathcal{L}$, the following set of measures
\begin{eqnarray}
\label{H4}
\mathcal{H}_4^{(\epsilon_4)}=\left\{d\tilde{\mu}_4\in \mathcal{L}:\quad 
x_b(\beta)|_{d\tilde{\mu}_4}=
{\frac{F^{\left(4\right)}_{b}}
{\left(F^{\left(2\right)}_{b}\right)^2}}|_{d\tilde{\mu}_4}<
e^{\frac{1}{2}a'-\epsilon_4},
\forall \beta\right\},
\end{eqnarray} 
with $0<\epsilon_4\ll a'/2$ 
and let us define a measure $d\mu_4$ in such a way that 
\begin{eqnarray}
\label{mu4}
\|d\mu_4-d\mu\|=
\inf\limits_{d\tilde{\mu}_4\in\mathcal{H}_4^{(\epsilon_4)}}\|
d\tilde{\mu}_4-d\mu\|.
\end{eqnarray} 
One can look at the measure $d\mu_4$ as an ``intermediate''
measure between the given measure $d\mu$ which, in general, does not
belong to the set $\mathcal{H}_4^{(\epsilon_4)}$, 
and the plus-minus delta measure $d\mu_{\pm}$
which satisfies the condition for $\mathcal{H}_4^{(\epsilon_4)}$. 
Note that $\mathcal{H}_4^{(\epsilon_4)}$ is dense set.
Equation (\ref{mu4}) implies
\begin{eqnarray}
\label{INEQ}
x_b(\beta)|_{d\mu_4}<e^{\frac{1}{2}a'}, \quad \forall \beta.
\end{eqnarray} 
In defining $\mathcal{H}_4^{(\epsilon_4)}$ we have introduced
a small parameter $\epsilon_4>0$ for having the above inequality strict,
furthermore, for reasons will become clear later, 
we have introduced the space $\mathcal{H}_4^{(\epsilon_4)}$
defined through the exponent $a'/2$ rather than $a'$.   
Therefore, for the modified measure $d\mu_4$, 
the power series (\ref{PEVEN4B}) converges 
and we can analyze the 
$\lim_{l_1,l_2\to\infty}~\lim_{N\to \infty}$ of
$B_N(l_1,l_2)$, by switching the limit of the series with the series of the limit. 
As a result we see that for the modified measure $d\mu_4$ we have 
\begin{eqnarray}\fl
\label{PEVEN4F}
\lim_{l_1,l_2 \to \infty} ~\lim_{N\to \infty}B_N(l_1,l_2)|_{d{\mu}_4}=
B_4, 
\quad \forall \beta,
\end{eqnarray} 
where $B_4=B_4(\beta)$ is a positive function of $\beta$ analytic
for any values of $\beta$.
Due to the fact that near the critical point only infinitely long
multi-paths contribute to the series and 
by noting that $C_N(l_1,l_2)=C_N(l_1)C_N(l_2)$, 
by using Eq. (\ref{PEVEN4F}) in Eq. (\ref{PEVEN4A}) we get
\begin{eqnarray}
\label{PEVEN4G}
\lim_{\beta\to\beta_4{-}}P^{(4)}_{N\to\infty}|_{d{\mu}_4}&=& 
\lim_{\beta\to\beta_4{-}}3 B_4
\left(P_{N\to\infty}\left(F_b^{(2)}\right)\right)^2|_{d{\mu}_4},
\end{eqnarray} 
where $\beta_4$ is the inverse critical temperature of the related Ising
model whose couplings, in terms of the high temperature expansion parameters
$\tanh(\beta J^{(I)})$ are substituted by the terms 
$\left(F_b^{(2)}\right)|_{d{\mu}_4}$.
This equation shows, in particular, that with the measure $d\mu_4$
the term $P^{(4)}$ is singular at the value $\beta_4$, \textit{i.e.}, the
inverse critical temperature $\beta_c^{\mathrm{(SG)}}$ given 
by the Eq. (\ref{mapp0g}) of the mapping.

Similarly, for $P^{(6)}$ from Eq. (\ref{PEVEN60}) we have
\begin{eqnarray}\fl
\label{PEVEN6}
P^{(6)}_N &=& 15\sum_{l_1,l_2,l_3,l^{(2)},l^{(3)}}^N 
C_N(l_1,l_2,l_3;l^{(2)},l^{(3)})
\left(F^{\left(2\right)}_{b}\right)^{l_1+l_2+l_3}
x_b^{l^{(2)}}y_b^{l^{(3)}},
\end{eqnarray} 
where $x_b$ is defined as in Eq. (\ref{PEVEN4x}), whereas $y_b$ is defined as
\begin{eqnarray}
\label{PEVEN6y}
y_b(\beta)\equiv 
\frac{F^{\left(6\right)}_{b}}{\left(F^{\left(2\right)}_{b}\right)^3}.
\end{eqnarray} 
Notice that, due to the H$\ddot{o}$lder inequality, 
as $x_b$, also $y_b\geq 1$ and it is equal to 1 only for the 
delta plus-minus measure. 
We recall that $C_N(l_1,l_2,l_3;l^{(2)},l^{(3)})$ represents
the number of triples of multi-paths of length $l_1,l_2,l_3$ which
share two to two and three to three $l^{(2)}$ and $l^{(3)}$ bonds, 
respectively.
Let us rewrite Eq. (\ref{PEVEN6}) as follows
\begin{eqnarray}
\label{PEVEN6A}
P^{(6)}_N &=& 15\sum_{l_1,l_2,l_3}^N C_N(l_1,l_2,l_3)B_N(l_1,l_2,l_3)
\left(F^{\left(2\right)}_{b}\right)^{l_1+l_2+l_3},
\end{eqnarray} 
where 
\begin{eqnarray}
\label{PEVEN6B}
B_N(l_1,l_2,l_3)&\equiv&
\frac{\sum_{l^{(2)},l^{(3)}}^{\bar{l^{(2)}},\bar{l^{(3)}}} 
C_N(l_1,l_2,l_3;l^{(2)},l^{(3)})x_b^{l^{(2)}} y_b^{l^{(3)}}}
{\sum_{l^{(2)},l^{(3)}}^{\bar{l^{(2)}},\bar{l^{(3)}}} 
C_N(l_1,l_2,l_3;l^{(2)},l^{(3)})} \nonumber \\
 &=& \sum_{l^{(2)},l^{(3)}}^{\bar{l^{(2)}},\bar{l^{(3)}}} 
\mathcal{P}_N(l_1,l_2,l_3;l^{(2)},l^{(3)}) 
x_b^{l^{(2)}}y_b^{l^{(3)}},
\end{eqnarray} 
\begin{eqnarray}
\label{PEVEN6l}
\bar{l^{(2)}} \equiv \mathrm{max}
\{\mathrm{min}(l_1,l_2);\mathrm{min}(l_1,l_3);\mathrm{min}(l_2,l_3)\}, 
\end{eqnarray} 
and
\begin{eqnarray}
\label{PEVEN6l'}
\bar{l^{(3)}} \equiv \mathrm{min}(l_1,l_2,l_3).
\end{eqnarray} 
Now, by using standard probability arguments as shown in Appendix B, 
it is easy to see that the condition (\ref{PROB7}) ensures also
\begin{eqnarray}
\label{PROB9}\fl
\mathcal{P}(l_1,l_2,l_3;l^{(2)},l^{(3)})
\leq (C')^2\left[
\frac{\left(l^{(2)}\right)^2+\left(l^{(3)}\right)^2}{2}
\right]^\frac{1}{2} e^{-\frac{1}{2}a'l^{(2)}-\frac{1}{2}a'l^{(3)}}.
\end{eqnarray} 
We find it convenient to observe that Eq. (\ref{PROB9}) in particular implies 
\begin{eqnarray}
\label{PROB9b}\fl
\mathcal{P}(l_1,l_2,l_3;l^{(2)},l^{(3)})<
(C')^2 \mathrm{max}\{l^{(2)},l^{(3)}\} \frac{4}{{(\frac{3}{2})!}^{\frac{1}{2}}}
e^{-\frac{1}{4}a'l^{(2)}-\frac{1}{4}a'l^{(3)}}.
\end{eqnarray} 
Let us introduce the following set of measures
\begin{eqnarray}\fl
\label{H6}
\mathcal{H}_6=\left\{d\tilde{\mu}_6\in \mathcal{L}:\quad 
\frac{F^{\left(4\right)}_{b}}
{\left(F^{\left(2\right)}_{b}\right)^2}|_{d\tilde{\mu}_6}<e^{\frac{1}{4}a'},
\frac{F^{\left(6\right)}_{b}}
{\left(F^{\left(2\right)}_{b}\right)^3}|_{d\tilde{\mu}_6}<e^{\frac{1}{4}a'},
\forall \beta \right\},
\end{eqnarray} 
and let us define a modified measure $d\mu_6$ in such a way that 
\begin{eqnarray}
\label{mu6}
\|d\mu_6-d\mu\|=
\inf\limits_{d\tilde{\mu}_6\in\mathcal{H}_6}\|d\tilde{\mu}_6-d\mu\|.
\end{eqnarray} 

From Eq. (\ref{mu6}) for the measure $d\mu_6$ we have 
that for any $\beta$,
$x_b(\beta)|_{d{\mu}_6}<e^{a'/4}$ and
$y_b(\beta)|_{d{\mu}_6}<e^{a'/4}$ so that, according to Eq. (\ref{PROB9b}),
for any $\beta$ 
the power series (\ref{PEVEN6B}) converges and,
as in the previous case, we can evaluate 
the $\lim_{N\to \infty}\lim_{l_1,l_2,l_3\to\infty}$ of
$B_N(l_1,l_2,l_3)$ by switching the limit of the series with the series
of the limit obtaining
\begin{eqnarray}
\label{PEVEN6F}
\lim_{l_1,l_2,l_3 \to \infty} 
\lim_{N\to \infty}B_N(l_1,l_2,l_3)|_{d{\mu}_6}=
B_6,
\end{eqnarray} 
where $B_6=B_6(\beta)$ is a positive function of $\beta$ analytic
for any values of $\beta$.
Finally, by noting that $C_N(l_1,l_2,l_3)=C_N(l_1)C_N(l_2)C_N(l_3)$ 
from Eq. (\ref{PEVEN6A}) we get
\begin{eqnarray}
\label{PEVEN6G}
\lim_{\beta\to\beta_6{-}}P^{(6)}_{N\to\infty}|_{d{\mu}_6}&=& 
\lim_{\beta\to\beta_6{-}}15 B_6
\left(P_{N\to\infty}\left(F_b^{(2)}\right)\right)^3|_{d{\mu}_6},
\end{eqnarray} 
where $\beta_6$ is the inverse critical temperature of the related Ising
model whose couplings, in terms of the high temperature expansion parameters
$\tanh(\beta J^{(I)})$, are substituted by the terms 
$\left(F_b^{(2)}\right)|_{d{\mu}_6}$.
This equation shows, in particular, that with the measure $d\mu_6$
the term $P^{(6)}$ is singular at the value $\beta_6$, \textit{i.e.}, the
inverse critical temperature $\beta_c^{\mathrm{(SG)}}$ given 
by the Eq. (\ref{mapp0g}) of the mapping.

Equation (\ref{PROB9}) can be generalized to any integer $n$ (see Appendix B) 
\begin{eqnarray}
\label{PROB10}\fl
\mathcal{P}(l_1,\ldots,l_{n};l^{(2)},\ldots,l^{(n)})
&\leq& (C')^n\left[
\frac{\left(l^{(2)}\right)^{n-1}+\ldots+\left(l^{(n)}\right)^{n-1}}{(n-1)!}
\right]^\frac{1}{n-1} \nonumber \\ 
&& \times e^{-\frac{1}{n-1}a'l^{(2)}-\ldots - \frac{1}{n-1}a'l^{(n)}}
\end{eqnarray} 
which, for $n> 1$, leads also to the generalization of Eq. (\ref{PROB9b})
\begin{eqnarray}
\label{PROB10B}\fl
\mathcal{P}(l_1,\ldots,l_{n};l^{(2)},\ldots,l^{(n)})
&<& (C')^n \mathrm{max}\{l^{(2)},\ldots,l^{(n)}\} 
\frac{n+1}{{(\frac{n}{2})!}^{\frac{1}{n-1}}}
e^{-\frac{1}{2n}a'l^{(2)}-\ldots - \frac{1}{2n}a'l^{(n)}}.
\end{eqnarray} 
Hence, for any $n>1$, by repeating the same argument followed for
$P^{(4)}_{N\to\infty}|_{d{\mu}_{4}}$ and
$P^{(6)}_{N\to\infty}|_{d{\mu}_{6}}$ we get 
\begin{eqnarray}
\label{PEVEN2nG}
\lim_{\beta\to\beta_{2n}{-}}P^{(2n)}_{N\to\infty}|_{d{\mu}_{2n}}&=& 
\lim_{\beta\to\beta_{2n}{-}}\frac{(2n)!}{2^n}B_{2n}
\left(P_{N\to\infty}\left(F_b^{(2)}\right)\right)^n|_{d{\mu}_{2n}},
\end{eqnarray} 
where: $B_{2n}=B_{2n}(\beta)$ is an analytic function of $\beta$ for any $\beta$,
$d{\mu}_{2n}$ is a modified measure defined as
\begin{eqnarray}
\label{mu2n}
\|d\mu_{2n}-d\mu\|=
\inf\limits_{d\tilde{\mu}_{2n}\in\mathcal{H}_{2n}}\|d\tilde{\mu}_{2n}-d\mu\|,
\end{eqnarray} 
with
\begin{eqnarray}\fl
\label{H2n}
\mathcal{H}_{2n}=\left\{d\tilde{\mu}_{2n}\in \mathcal{L}:\quad 
\frac{F^{\left(4\right)}_{b}}
{\left(F^{\left(2\right)}_{b}\right)^2}|_{d\tilde{\mu}_{2n}},
\ldots, \frac{F^{\left(2n\right)}_{b}}
{\left(F^{\left(2\right)}_{b}\right)^n}|_{d\tilde{\mu}_{2n}}<e^{\frac{1}{2n}a'}.
\forall \beta \right\},
\end{eqnarray} 
and $\beta_{2n}$ is the inverse critical temperature of the related Ising
model whose couplings, in terms of the high temperature expansion parameters
$\tanh(\beta J^{(I)})$, are substituted by the terms 
$\left(F_b^{(2)}\right)|_{d{\mu}_{2n}}$.
This equation shows, in particular, that with the measure $d\mu_{2n}$
the term $P^{(2n)}$ is singular at the value $\beta_{2n}$, \textit{i.e.}, the
inverse critical temperature $\beta_c^{\mathrm{(SG)}}$ given 
by the Eq. (\ref{mapp0g}) of the mapping.

Let us define $\beta_0=\beta_c^{\mathrm{(SG)}}$, the solution of Eq. (\ref{mapp0g}) of
the mapping with the given measure $d\mu$. 
We have now to calculate the free energy density term $\varphi$, 
by using the replica trick relation (\ref{phi1}) or, equivalently 
\begin{eqnarray}
\label{phi1EVEN}
\varphi = \lim_{n \rightarrow 0}\lim_{N \rightarrow \infty} 
\frac{P^{(2n)}-1}{2nN}.
\end{eqnarray} 
From the structure of the generic set $\mathcal{H}_{2n}$ it is
immediate to recognize that $\mathcal{H}_{4}\supset\mathcal{H}_{6}\supset
\ldots\supset\mathcal{H}_{2n}$ and that for $n\to\infty$ 
$\mathcal{H}_{2n}\to\{d\mu_{\pm}\}$. However we are interested in the opposite
limit $n\to 0^+$. 
Let us observe that Eqs. (\ref{PROB10B}-\ref{H2n}) allow to be analytically
continued to any real $n\geq 0$, with the pre-factor
$C^n(n+1)/((n/2)!^{1/(n-1)}\to 1$ as $n\to 0$.
In this limit, the analytic continuation of the constrains
in Eq. (\ref{H2n}) brings to $\mathcal{H}_{2n}\to\mathcal{L}$ 
and Eq. (\ref{mu2n}) gives $d\mu_{2n}\to d\mu$ with respect to the functional
distance $\|\cdot\|$. 
Therefore, for Eq. (\ref{phi1EVEN}) we are free to calculate
the limit as 
\begin{eqnarray}
\label{phi1EVEN2}
\varphi = \lim_{n \rightarrow 0}\lim_{N \rightarrow \infty} 
\frac{P^{(2n)}_N-1}{2nN}=
\lim_{n \rightarrow 0}\lim_{N \rightarrow \infty} 
\frac{P_N^{(2n)}|_{d\mu_{2n}}-1}{2nN}
\end{eqnarray} 
and from Eq. (\ref{PEVEN2nG}), and by using 
$\lim_{\beta\to\beta_0}f(\beta)=\lim_{n\to 0}f(\beta_{2n})$, we finally get:
\begin{eqnarray}
\label{phi1EVEN2}
\lim_{\beta\to\beta_0{-}} \varphi = 
\lim_{\beta\to\beta_0{-}}  \lim_{N \rightarrow \infty}
\frac{\log\left(P_N\left(F_b^{(2)}\right)\right)}{N} =
\lim_{\beta\to\beta_0{-}}\frac{1}{2}\varphi_I\left(F_b^{(2)}\right).
\end{eqnarray} 

We stress that within the replica trick this proof is exact.
With Eqs. (\ref{PROB10B}-\ref{H2n}) we have indeed found that 
there exists a succession of spaces $\mathcal{H}_{2n}$ where
Eq. (\ref{PEVEN2nG}) holds with $B_{2n}$ finite and that this succession
can be analytically continued to any real $n\geq 0$.
It is important to note here that such a situation does not apply
in a finite $D$-dimensional hypercube lattice. In fact, in this case
the probability $\mathcal{P}(l_1,l_2;l)$ behaves
in a completely different manner and cannot satisfy the condition 
(\ref{PROB2}).
More precisely, in these ``particular'' graphs, for $D>1$, even if $C_N(l)$
still continues to have an exponential growth in $l$, $C_N(l)\sim (\bar{k}-1)^l$, 
with $\bar{k}-1=2D-1$, the probability $\mathcal{P}(l_1,l_2;l)$ 
remains concentrated 
to values of $l$ near to the given lengths $l_1$ and $l_2$.
We can easily understand this last statement by looking at the case $D=1$.
Even tough in this case $\bar{k}=2$, this example turns out to be quite 
instructive. In fact, in one dimension, for $\mathcal{P}(l_1,l_2;l)$ we have 
exactly
\begin{eqnarray}
\label{PPP2}
\mathcal{P}(l_1,l_2;l)=\left\{
\begin{array}{c}
1, \quad l=\mathrm{min}\{l_1,l_2\},\\
0, \quad \mathrm{otherwise}.
\end{array}
\right.
\end{eqnarray}
With such a probability, Eq. (\ref{PEVEN4B}) 
gives $B_N(l_1,l_2)=x_b^{\mathrm{min}\{l_1,l_2\}}$
so that for $l_1,l_2\to\infty$ $B_N(l_1,l_2)$ diverges 
and Eq. (\ref{PEVEN4G}) does not hold.
For $D>1$ and finite, in general, it is very difficult to calculate 
the probability $\mathcal{P}(l_1,l_2;l)$, but, roughly speaking, 
a similar behavior is expected as well \cite{Filippo}.

\subsection{Proof for any measure}
For a generic measure, \textit{i.e.} also non symmetric, we have $F_b^{(m)}\neq 0$
also for $m$ odd so that for calculating $P^{(2n)}$, besides the terms obtained
in the previous section, we have to add the contributions involving all
the possible odd overlaps due to $m$-overlaps with $m\geq 1$.
Let us define $\beta_{2n}$ as
\begin{eqnarray}
\label{PEVENGG}
\beta_{2n}=\mathrm{min}\{\beta^{\mathrm{(SG)}}_{2n},\beta^{\mathrm{(F/AF)}}_{2n}\},
\end{eqnarray} 
where $\beta^{(\mathrm{SG})}_{2n}$ and $\beta^{(\mathrm{F/AF})}_{2n}$ 
are the inverse critical temperatures of 
of the related Ising model whose couplings, in terms of the high temperature 
expansion parameters $\tanh(\beta J^{(I)})$ are substituted by the terms 
$\left(F_b^{(2)}\right)|_{d{\mu}_{4n}}$ and $\left(F_b^{(1)}\right)|_{d{\mu}_{4n}}$,
respectively.
On the same line of the Step 2 of the proof, we have that the singular behavior 
of the terms $P^{(2n)}$ is described by
\begin{eqnarray}\fl
\label{PEVEN2GG}
\lim_{\beta\to\beta_2{-}}P^{(2)}_{N\to\infty}|_{d{\mu_4}}&=& 
\lim_{\beta\to\beta_2{-}}
\left(P_{N\to\infty}\left(F_b^{(2)}\right)\right)|_{d{\mu_4}}+ 
\nonumber \\ 
&&\lim_{\beta\to\beta_2{-}}3 B_4
\left(P_{N\to\infty}\left(F_b^{(1)}\right)\right)^2|_{d{\mu}_4},
\end{eqnarray} 
\begin{eqnarray}\fl
\label{PEVEN4GG}
\lim_{\beta\to\beta_4{-}}P^{(4)}_{N\to\infty}|_{d{\mu}_8}&=& 
\lim_{\beta\to\beta_4{-}}3 B_4
\left(P_{N\to\infty}\left(F_b^{(2)}\right)\right)^2|_{d{\mu}_8}+
\nonumber \\
&&\lim_{\beta\to\beta_4{-}} B_8
\left(P_{N\to\infty}\left(F_b^{(1)}\right)\right)^4|_{d{\mu}_8},
\end{eqnarray} 
\begin{eqnarray}\fl
\label{PEVEN2nGG}
\lim_{\beta\to\beta_{2n}{-}}P^{(2n)}_{N\to\infty}|_{d{\mu}_{4n}}&=& 
\lim_{\beta\to\beta_{2n}{-}}\frac{(2n)!}{2^n}B_{2n}
\left(P_{N\to\infty}\left(F_b^{(2)}\right)\right)^n|_{d{\mu}_{4n}}+
\nonumber \\
&&\lim_{\beta\to\beta_{2n}{-}}B_{4n}
\left(P_{N\to\infty}\left(F_b^{(1)}\right)\right)^{2n}|_{d{\mu}_{4n}},
\end{eqnarray} 
where we have made use of the fact that $\mathcal{H}_{2n}\supset\mathcal{H}_{2n+2}$.
By using Eq. (\ref{PEVEN2nGG}), $\mathcal{H}_{2n}\to \mathcal{L}$ as $n\to 0^+$, 
and Eq. (\ref{phi1EVEN}) we see that the upper critical
surface is given by Eq. (\ref{mappg}). 
Finally, the generalization to an arbitrary measure $d\mu_b$ which depends also
on the bond $b\in\Gamma$ is straightforward: Eq. (\ref{PEVEN2nGG}) has to be
substituted with an analogous equation in which we have simply to replace
$d\mu_{4n}$, $F_b^{(2)}$ and $F_b^{(1)}$ with the corresponding vectors 
$\{d\mu_b\}_{4n}$, $\{F_b^{(2)}\}$ and $\{F_b^{(1)}\}$, respectively.

\section{The case of the measure $\mu_{\pm}$}
In the previous section, we have proved the mapping
in two steps, along the subsections \ref{First} and \ref{Second}.
Unlike the second step, in the first step,
the measure $d\mu$ does not play any role. However,
from Eq. (\ref{PEVEN60Gpm}) we see that the measure $d\mu_\pm$ results to be a very
special measure; in fact this equation says that if $d\mu=d\mu_{\pm}$,
the mapping turns out to be exact even in finite dimension.
In particular, in $D=2$ dimensions a phase transition should exist.
This result seems to be in contradiction with the known fact,
from numerical simulations, that in $D=2$ dimensions 
there is no phase transition. The paradox is explained by looking at
the Step 2 of the proof. From this part of the proof it becomes
clear that within the set of all the possible measures, the measure
$d\mu_{\pm}$ constitutes a singular measure, being the unique
for which Eq. (\ref{PEVEN60Gpm}) can be satisfied and, as soon as a measure $d\mu$
is even infinitesimally different from the measure $d\mu_{\pm}$,
Eq. (\ref{PEVEN60Gpm}) cannot be satisfied; the given measure $d\mu$ 
may instead satisfy Eq. (\ref{PEVEN60G})
but only in infinite dimensions. Therefore, within the set of all the measures,
the measure $d\mu_{\pm}$ represents a singular measure for which there is
an unstable phase transition with no physical counterpart. 
We recall that even in a numerical
experiment it is impossible to represent exactly the singular measure
$d\mu_{\pm}$. In fact, one can try to reproduce numerically such a distribution
of bonds approximately by a smooth modification, but not exactly.

\section{Nishimori law}\label{Nishi}
Let us consider an arbitrary measure $d\mu$ independent on the bond $b$. 
From the equations of the mapping (\ref{PROB2mapp}-\ref{mapp01g}) 
we see that it may exist 
a tricritical point
$\beta_{ct}=\beta_c^{(\mathrm{SG})}=\beta_c^{(\mathrm{F})}$ 
where the phases F, P and SG meet given by 
\begin{eqnarray}
\label{NISHI}
\int d\mu \tanh^2(\beta_{ct} J_b)=\int d\mu \tanh (\beta_{ct} J_b).
\end{eqnarray} 
In the particular case of the measure
\begin{eqnarray}
\label{NISHI2}
\frac{d\mu(J_b)}{dJ_b}=p\delta(J_b-J)+(1-p)\delta(J_b+J), \qquad 0\leq p\leq 1,
\end{eqnarray} 
Eq. (\ref{NISHI}) gives
\begin{eqnarray}
\label{NISHI3}
\tanh (\beta_{ct} J)=2p-1.
\end{eqnarray} 
From Eq. (\ref{NISHI3}) we see that, as it must be, for any choice of $p$,
the multicritical point F-P-SG belongs to the Nishimori line \cite{Nishimori}.
Equation (\ref{NISHI}) can be seen as a generalization of the Nishimori line
to any measure. 

Let us come back to the measure (\ref{NISHI2}). 
The Nishimori theorems say also that, for any $p$, the internal energy along the
Nishimori line $\tanh(\beta J)=2p-1$ 
is given by $E=-(NkJ/2)\tanh(\beta J)$ or, equivalently, the free energy 
is given by $-\beta F=(Nk/2) \log(2\cosh(\beta J))$, where $k$ is the mean connectivity,
or coordination number. According to our definition of $\varphi$, Eq. (\ref{phi0}),
this implies that $\varphi=0$ along the Nishimori line. In the framework of
the mapping this result can be checked from the fact that along the P-F and the
P-SG lines $\varphi$ is given by $\varphi_I((2p-1)\tanh(\beta_{c}J))$ 
and $\varphi_I(\tanh^2(\beta_{c}J))/2$, respectively, so that on the multicritical
point $\beta_{ct}$ it must be 
$\varphi_I((2p-1)\tanh(\beta_{c}J))=\varphi_I(\tanh^2(\beta_{c}J))/2$. 
From Eq. (\ref{NISHI3}) we see that the only solution of this 
last equation can be only $\varphi=\varphi_I\equiv 0$ along all the upper critical line
and, for continuity, in all the P region. By repeating this argument in the
general case for an arbitrary measure $d\mu$ independent on $b$, 
and by using Eq. (\ref{NISHI}), we get
\begin{eqnarray}
\label{NISHI6}
\varphi=\varphi_I\equiv 0, \quad \mathrm{in~the~P~region},
\end{eqnarray} 
or equivalently
\begin{eqnarray}\fl
\label{NISHI7}
E=-\frac{Nk}{2}\int d\mu(J_b) J_b \tanh(\beta J_b), \quad
E_I=-\frac{Nk}{2}J^{(I)} \tanh(\beta J^{(I)}), \quad \mathrm{in~P}.
\end{eqnarray} 
Notice that this result holds also for the related Ising model.

A further generalization can be also considered when
the measure depends also on the bond $b$. From Eqs. (\ref{PROB2mapp}-\ref{mapp01g})
we see that the multicritical point 
$\beta_{ct}=\beta_c^{(\mathrm{SG})}=\beta_c^{(\mathrm{F/AF})}$ 
must satisfy the two following - possibly vectorial - equations
\begin{eqnarray}
\label{NISHI4}
G_I\left(\left\{\int d\mu_b(J_b) \tanh (\beta_{ct} J_b)\right\}\right)=0, 
\end{eqnarray} 
and
\begin{eqnarray}
\label{NISHI5}
G_I\left(\left\{\int d\mu_b(J_b) \tanh^2 (\beta_{ct} J_b)\right\}\right)=0. 
\end{eqnarray} 
These equations may be regarded as the most general formulation of the Nishimori surface
for models defined over graphs infinite dimensional in the broad
sense. Here we use the term Nishimori surface, just to indicate a
surface passing through the multicritical points 
(or even multicritical surfaces) and having the
following property: as we change continously the parameters of the measures $\{d\mu_b\}$
as to approach a single measure $d\mu$ no longer dependent on the bond $b$, we 
approach the Nishimori line (\ref{NISHI}) for which 
Eqs. (\ref{NISHI6}) and (\ref{NISHI7}) hold.

We claim that Eq. (\ref{NISHI6}) (or (\ref{NISHI7})) 
is the feature characterizing the models defined over
graphs infinite dimensional in the broad sense: $\varphi$ (or $\varphi_I$) 
is zero not only along the Nishimori line, but over the whole P region.

Note however that there can be exceptions to the above rule for the graphs where
the density free energy $f$ (and then also $\varphi_I$) does not exist 
in the proper sense.
As mentioned in Sec. 4, this may happen in considering models over Husimi
trees. Like in the Bethe lattices,
in the Husimi trees, the thermodynamic limit of the density free energy does not exist.
In these kind of models in fact, what is sensible is only the physics of the central spin:
its magnetization and correlation functions obtained with recursive relations \cite{MOIII}.
From these quantities one can then define the free energy 
\textit{a posteriori} as done in \cite{Baxter}.
It is possible to check that in a Bethe lattice one has however $\varphi_I\equiv 0$ in the
P region. However, that does not happen in a Husimi tree (we have seen this for the model considered
in \cite{MOIII}). Note that this is not in contradiction with the theorem of Nishimori.
In fact, the Nishimori theorem, as well as our mapping, concerns, rigorously speaking, only 
``regular'' models for which the free energy has some thermodynamic limit 
(see Eq. (6) of Ref. \cite{Nishimori}). The models defined over Bethe lattices and Husimi trees
therefore may, in general, 
not satisfy the Nishimori theorems. The fact that over Bethe lattices
they still satisfy the Nishimori theorems is accidentally due to the tree-like structure of
the Bethe lattices. 

\section{Conclusions}
We have deepened concepts and claims already mentioned and
used in the Refs. \cite{MOI} and \cite{MOII} and we have 
provided a complete proof of the mapping for general graphs. 
The condition for the mapping to become exact for these graphs
is not exactly what claimed in \cite{MOI}, where we simply required 
a decay of $p(l)$ to 0 for $l$ going to infinity (such a requirement
is in fact only a necessary condition if $p(l)$ is a probability).
The mapping becomes exact whenever $p(l)$ decays at least exponentially
or, for the most general case (even if the limit $p(l)$ does not exist),
when Eq. (\ref{PROB2}) is satisfied. 
As Eq. (\ref{PROB2}) is satisfied we say that the graph is
infinite dimensional in the broad sense. The use of this expression
is motivated by the fact Eq. (\ref{PROB2}) implies the infinite dimensionality 
in the traditional sense (see Eqs. (\ref{PROB4}) and \ref{PROBDIS}) but not vice-versa.
In fact, this non equivalence represents an interesting point, since
there exist graphs where the tree-like approximation cannot be applied
but, nevertheless, they are infinite dimensional in the broad sense.
Also from this observation, it should be clear that the proof of the mapping,
far from being simple, is not based on some local analysis of the graph,
but on the key requirement expressed by Eq. (\ref{PROB2}), which represents 
a global information; a crucial feature in spin glass models.
Note that, along the proof, no ansatz, such as the replica symmetric one, has been used.

The powerful of the mapping in respect of its simplicity and generality
has still to be explored. Applications of the mapping to random graphs
not yet considered in Ref. \cite{MOII} and its extensions to include
graphs with constrains (complex networks) and others non Ising models
are the object of present and future investigations.

\section*{Acknowledgments}
This work was supported by the FCT (Portugal) grant SFRH/BPD/24214/2005. 
I am grateful to F. Mukhamedov, F. Cesi and F. Ricci-Tersenghi 
for many useful discussions.

\appendix

\addcontentsline{toc}{section}{A sufficient condition}
\section{A sufficient condition}
\setcounter{section}{1}

The mapping becomes exact under the condition
(\ref{PROB7}) for $\mathcal{P}(l_1,l_2;l)$, the
probability that two infinitely long - multi-paths - 
arbitrarily chosen overlap for $l$ bonds. The probability $\mathcal{P}(l_1,l_2;l)$
is however an awkward quantity of not direct access.
We are more interested in using the probability $p(l_1,l_2;l)$ introduced in Sec. 3 
that two infinitely long - simple - paths overlap for $l$ bonds. 
In fact, unlike $\mathcal{P}(l_1,l_2;l)$, $p(l_1,l_2;l)$ concerns only simple
connected paths (mono-component) and it is therefore a suitable quantity of
easier practical access.
We want now to find a sufficient condition for 
Eqs. (\ref{PROB7}) to be satisfied in terms of $p(l_1,l_2;l)$.
We will prove that the exponential decaying upper bound for $p(l_1,l_2;l)$ implies
that of $\mathcal{P}(l_1,l_2;l)$ with a little but finite diminishing of the exponent.

Given $l_1$, $l_2$ and $l\leq \mathrm{min}\{l_1,l_2\}$,
let $\mathcal{P}_N(l_1,l_2;l|k)$ be the conditional probability that two
multi-paths of length $l_1$ and $l_2$ overlap
for $l$ bonds along $k$ common portions of $k$ simple multi-paths 
of the first and second multi-path. The $k$ portions
belong to a common multi-path of length $l$.

For definition of $\mathcal{P}_N(l_1,l_2;l)$ we have
\begin{eqnarray}
\label{PP}
\mathcal{P}_N(l_1,l_2;l)=\sum_{k}^{l}
\mathcal{P}_N(l_1,l_2;l|k)\mathcal{P}_N(l;k),
\end{eqnarray} 
where $\mathcal{P}_N(l;k)$ is the probability that
the common multi-path of length $l$ is constituted by $k$ components (or
simple paths).

Let us indicate with $\xi_1,\ldots,\xi_{k}$ 
the numbers of two to two shared bonds along  
the $k$ common portions of the $k$ simple multi-paths of 
the two multi-paths.
Looking at the $\xi$'s as random variables, we have
\begin{eqnarray}
\label{PP1}
\mathcal{P}_N(l_1,l_2;l|k)=\sum_{x_1+\ldots +x_{k}=l}
\mathcal{P}_N(l_1,l_2;\xi_1=x_1,\ldots,\xi_k=x_{k}),
\end{eqnarray} 
where 
$\mathcal{P}_N(l_1,l_2;\xi_1=x_1,\ldots,\xi_{k}=x_{k})$ 
is the probability that two multi-paths of lengths $l_1$ and $l_2$
overlap two to two for $x_1,\ldots,x_{k}$ bonds along the 
$k$ common regions of the $k$ simple paths.
Note that, for definition of a path, the $k$ components cannot share common bond each
other. In fact we can say more.
We are interested in the limit $\lim_{l\to\infty}\lim_{l_1,l_2\to\infty}$.
In words this means that we keep $l$ fixed and look at the situation in which
the total lengths $l_1$ and $l_2$ of the two multi-paths become larger and
larger. In this limit the $k$ portions become farer and farer islands of
fixed size. Therefore, in the limit of interest, 
we can approximate the random variables
$\xi_1,\ldots,\xi_{k}$ as independent so that the probability
$\mathcal{P}_N(l_1,l_2;\xi_1=x_1,\ldots,\xi_{k}=x_{k})$ 
factorizes as
\begin{eqnarray}
\label{PP2}
\mathcal{P}_N(l_1,l_2;\xi_1=x_1,\ldots,\xi_{k}=x_{k})=
p_N(l_1',l_2';x_1)\cdots p_N(l_1',l_2';x_{k}),
\end{eqnarray} 
where $l_1'$ and $l_2'$ are rescaled lengths of the order of
$l_1'=\mathop{O}(l_1/k)$ and 
$l_2'=\mathop{O}(l_2/k)$. We observe that in the
limit $l_1,l_2\to \infty$ we have also $l_1',l_2'\to \infty$.

In the limit $l\to\infty$, we are allowed, for the sake of simplicity,
to take the variable $l$ in the continuum. Note that in such a case
we have to consider the following substitution rules
\begin{eqnarray}
\label{PP3}
\sum_l \Rightarrow \int dl \\
\lim_{l_1,l_2\to\infty}C_{l_1,l_2}=\frac{e^a-1}{e^a} \Rightarrow 
\lim_{l_1,l_2\to\infty}C_{l_1,l_2}=a \\
\sum_{x_1+\ldots +x_{k}=l} \Rightarrow \int dx_1\cdots dx_k\delta({x_1+\ldots +x_{k}-l}).
\end{eqnarray} 

The great simplification in considering continuum variables stands in the last
rule. In fact, we are able to calculate easily the above multi integral as
\begin{eqnarray}
\label{PP4}
\int_{0}^l dx_1\cdots dx_k\delta({x_1+\ldots +x_{k}-l})=\frac{l^{k-1}}{(k-1)!}.
\end{eqnarray} 

If we assume now the exponential decay (\ref{PROB2}) for $p(l_1,l_2;l)$, 
by using Eqs. (\ref{PP2}) and (\ref{PP4}), Eq. (\ref{PP1}) becomes
\begin{eqnarray}
\label{PP5}
\mathcal{P}_N(l_1,l_2;l|k)\leq C^k e^{-al}\frac{l^{k-1}}{(k-1)!}.
\end{eqnarray} 
Taking into account the second of the Eqs. (\ref{PP3}), 
in the limit $l_1,l_2\to \infty$ the normalization of the r.h.s. of
Eq. (\ref{PP5}) can be checked immediately by using 
\begin{eqnarray}
\label{PP6}
\int_0^{\infty} e^{-al}l^n =\frac{n!}{a^n}.
\end{eqnarray} 

The evaluation of the probability $\mathcal{P}_N(l;k)$ for having $k$
components (islands) of total length $l$, can be easily
obtained along the same line with the use of Eq. (\ref{PP4}).
In fact, since, given $l$, all the components $x_1,\ldots,x_k$ are
equiprobable, the only constrain for the random variable $k$ is that be
$x_1+\ldots +x_k=l$. So that, in the continuum for $l$, from Eq. (\ref{PP4}) we get
\begin{eqnarray}
\label{PP7}
\mathcal{P}_N(l;k)=\frac{l^{k-1}}{(k-1)!}e^{-l},
\end{eqnarray} 
where the factor $\exp(-l)$ comes from the normalization constant. 
Equation (\ref{PP7}) could be derived also by observing that, given $l$, 
$k$ describes the number of jumps inside the interval $[0,l]$ 
of a homogeneous jump process which, in correspondence of any bond $b$ of length 1,
may or may not to jump to another island,
so that $k$ is distributed according to a Poisson distribution with rate $\rho=1$.

Now, from Eqs. (\ref{PP}), (\ref{PP5}) and (\ref{PP7}), by passing in the 
continuum also for $k$ and by using the Stirling approximation we have
\begin{eqnarray}
\label{PP8}
\mathcal{P}_N(l_1,l_2;l)\leq\int_{0}^{l} dk e^{\psi(k)}\frac{1}{2\pi (k-1)},
\end{eqnarray} 
where we have introduced the non smooth part of the integrand
\begin{eqnarray}\fl
\label{PP9}
\psi(k)=-al-l+k\log(C)+2(k-1)\log(l)-2[(k-1)\log(k-1)-(k-1)].
\end{eqnarray} 
By a saddle point calculation in $k$ we get, 
$\partial_k \psi(k)=0$ for $k^{\mathrm{sp}}=l\sqrt{C}$, and using
$\partial_k^2 \psi(k)|_{k^{\mathrm{sp}}}=-2/(l\sqrt{C})$, for 
large $l$ we arrive at
\begin{eqnarray}
\label{PP10}
\mathcal{P}_N(l_1,l_2;l)\leq e^{-l(a-2\sqrt{C}+1)}\frac{1}{(2\pi)^{1/2}C^{1/4}\sqrt{l}}.
\end{eqnarray}  
Finally, by using $C=a$ we have
\begin{eqnarray}
\label{PP11}
\mathcal{P}_N(l_1,l_2;l)\leq e^{-l(a-2\sqrt{a}+1)}\frac{1}{(2\pi)^{1/2}a^{1/4}\sqrt{l}}.
\end{eqnarray}  

Since $a'\equiv a-2\sqrt{a}+1>0$ for any real $a$, we have proved that an exponential
decay in $l$ for $p_N(l_1,l_2;l)$ implies always an exponential decay for
$\mathcal{P}_N(l_1,l_2;l)$. It is interesting to observe that the rate
of decay of the latter is diminished by the term $2\sqrt{a}-1$, which is
positive for $a>1/4$.

\addcontentsline{toc}{section}{Inequalities}
\section{Inequalities}

Equation (\ref{PROB9}) can be proved as follows.
If $\mathcal{P}$ is some probability acting on sets $A_1,\ldots,A_n$, one has
\begin{eqnarray}
\label{B1}
\mathcal{P}(A_1\cap\ldots\cap A_n )\leq 
\left(\mathcal{P}(A_1)\cdots\mathcal{P}(A_n)\right)^{\frac{1}{n}},
\end{eqnarray} 
so that for $\mathcal{P}(l_1,l_2,l_3;l^{(2)},l^{(3)})$ we have
\begin{eqnarray}
\label{B2}\fl
\mathcal{P}(l_1,l_2,l_3;l^{(2)},l^{(3)})
\leq \left(\mathcal{P}^{(2)}(l_1,l_2,l_3;l^{(2)})
\mathcal{P}^{(3)}(l_1,l_2,l_3;l^{(3)})\right)^{\frac{1}{2}},
\end{eqnarray} 
where, for $m\leq n$, we have introduced the probabilities
$\mathcal{P}^{(m)}(l_1,\ldots,l_n;l^{(m)})$ for having a total of
$l^{(m)}$ $m$-overlaps among the $n$ paths.
So, the first factor in the rhs of Eq. (\ref{B2}) represents the probability for
having a 2-overlap $l^{(2)}$ among a triplet of multi-paths of lengths
$l_1,l_2,l_3$, whereas the second factor represents the probability for
having a 3-overlap $l^{(3)}$ among the triplet. 

On the same line of what we have seen in the Appendix A,
we observe that, if $\xi_{12}$, $\xi_{13}$ and $\xi_{23}$, are
the random variables associated with the total 2-overlap between the
multi-paths 1 and 2, the multi-paths 1 and 3, and the multi-paths 2 and 3, respectively, 
we can decompose the probability $\mathcal{P}(l_1,l_2,l_3;l^{(2)})$ as follows 
\begin{eqnarray}\fl
\label{B3}
\mathcal{P}_N^{(2)}(l_1,l_2,l_3;l^{(2)})=\sum_{x_{12}+x_{13}+x_{23}=l^{(2)}}
\mathcal{P}_N^{(2)}(l_1,l_2,l_3;\xi_{12}=x_{12},\xi_{13}=x_{13},\xi_{23}=x_{23}).
\end{eqnarray} 
On the other hand, in the limit $\lim_{l\to\infty}\lim_{l_1,l_2,l_3\to\infty}$,
we can apply the same argument we have used in the Appendix A: the three
families of regions between the multi-paths 1 and 2, 1 and 3, and 2 and 3,
become infinitely far in this limit, so that the respective random variables
$\xi_{12}$, $\xi_{13}$ and $\xi_{23}$ become independent and distributed
according to the exponential decay (\ref{PROB2}).
Hence, after using Eq. (\ref{PP4}) we are left with 
\begin{eqnarray}
\label{B4}\fl
\mathcal{P}^{(2)}(l_1,l_2,l_3;l^{(2)})\leq 
(C')^3 e^{-a'l^{(2)}}\frac{\left(l^{(2)}\right)^2}{2},
\end{eqnarray} 
where we have used Eq. (\ref{PROB7}).

Concerning the second factor, we simply observe that $l^{(3)}$ stands for a
3-overlap, therefore, by using the elementary inequality
\begin{eqnarray}
\label{B5}
\mathcal{P}(A\cap B \cap C )\leq \mathcal{P}(A\cap B),
\end{eqnarray} 
we get
\begin{eqnarray}
\label{B6}\fl
\mathcal{P}^{(3)}(l_1,l_2,l_3;l^{(3)})\leq
\mathcal{P}^{(2)}(l_1,l_2,l_3;l^{(3)}).
\end{eqnarray} 
By using Eqs. (\ref{B2}-\ref{B5})) Eq. (\ref{PROB9}) is proved.

Similarly, for the general case, the analogous of Eq. (\ref{B4}) becomes
\begin{eqnarray}
\label{B7}\fl
\mathcal{P}^{(2)}(l_1,\ldots,l_n;l^{(2)})\leq 
(C')^n e^{-a'l^{(2)}}\frac{\left(l^{(2)}\right)^{n-1}}{(n-1)!},
\end{eqnarray} 
and, for any $2\leq m\leq n$, the analogous of Eq. (\ref{B6}) becomes
\begin{eqnarray}
\label{B8}\fl
\mathcal{P}^{(m)}(l_1,\ldots,l_n;l^{(m)})\leq
\mathcal{P}^{(2)}(l_1,\ldots,l_n;l'^{(2)}).
\end{eqnarray} 
By using Eqs. (\ref{B2}), (\ref{B7})) and (\ref{B8}), Eq. (\ref{PROB10}) is proved.


\begin{thebibliography}{99}
\bibitem{NOTA} More precisely, given a graph, the Bethe-Peierls approximation
is exact when the corresponding factor graph is a tree. When in the factor
graph there are loops, the Bethe-Peierls approach gives good results provided
the length of the loops is sufficiently large (for example when the length
scales as $\mathop{O}(\log N)$), 
whereas, in the presence of short loops, corrections to the Bethe-Peierls approach
may be important. Systematic corrections to the Bethe-Peierls approximation
have been recently rigorously established in \cite{PS} and \cite{CC}.
\bibitem{MOIII} M. Ostilli, F. Mukhamedov, J. F. Mendes, cond-math/0611654. In this manuscript
we provide a mere application of the equation of the mapping regardless of the fact that,
as stressed, the free energy for models defined over a Husimi tree, does not exist and
the mapping here, rigorously speaking, cannot be applied.
\end{thebibliography}
\end{document}